\documentclass[
reprint,
nofootinbib,
 amsmath,amssymb,
 aps,
prd,
floatfix,
]{revtex4-2}

\usepackage{tikz}
\usepackage{hhline}
\usepackage{graphicx}
\usepackage{dcolumn}
\usepackage{bm}
\usepackage[caption=false]{subfig}
\usepackage{makecell}
\usepackage{placeins}
\usepackage{varwidth}
\usepackage[breaklinks,colorlinks]{hyperref}
\usepackage{booktabs}
\hypersetup{%
,urlcolor=blue
,citecolor=blue
,linkcolor=blue
}
\newcommand{\half}{{\frac{1}{2}}}
\newcommand{\quart}{{\frac{1}{4}}}

\begin{document}

\preprint{APS/123-QED}

\title{Kinky vortons in the 2HDM}

\author{Richard A. Battye}
    \email{richard.battye@manchester.ac.uk}
\author{Steven J. Cotterill}%
    \email{steven.cotterill@manchester.ac.uk}
\author{Adam K. Thomasson}%
    \email{adam.thomasson@manchester.ac.uk}

\affiliation{Department of Physics and Astronomy, The University of Manchester, Manchester, U.K.}

\date{\today}

\begin{abstract}
We construct and analyse two-dimensional, current-carrying ring solutions, known as kinky vortons, in the $\mathbb{Z}_2$-symmetric global two-Higgs-doublet model (2HDM). We demonstrate the existence of multiple dynamically stable configurations that persist under non-axially symmetric perturbations. These solutions are described with high accuracy by the thin string approximation and elastic string formalism, which correctly capture both their equilibrium radii and dynamical oscillation frequencies. Kinky vortons in the $\mathbb{Z}_2$-symmetric theory establish the viability of vorton solutions in a phenomenologically motivated extension of the Standard Model, and should provide a computationally tractable proxy for vortons in the $U(1)$-symmetric 2HDM. In addition, we identify a composite domain wall configuration in which localized condensates are supported on secondary domain walls existing on a $\mathbb{Z}_2$ wall, suggesting a mechanism by which kinky-vorton-like defects could arise in a three dimensional setting.
\end{abstract}

\maketitle


\section{Introduction}\label{sec:Intro}
Cosmic strings are line-like topological defects that may have formed during symmetry-breaking phase transitions in the early universe \cite{Vilenkin278400}. Witten \cite{Witten:1984eb} showed that cosmic strings can support persistent currents if the string-forming field couples to an additional complex scalar field. Specifically, he studied a $U(1) \times U(1)$ model where one symmetry is broken and the other remains unbroken in the vacuum. Rotating closed loops of superconducting string, known as vortons, supported by angular momentum, charge, and current, were first suggested by Davis and Shellard \cite{DAVIS1989209}.

The dynamics and stability of vortons remained open questions long after their formulation, due to the computational challenges associated with simulating them and the limited exploration of parameter space \cite{Lemperiere:2003yt, Battye:2008mm, Garaud_2013}. Because vortons of cosmological relevance are expected to be much larger than the string width, it is believed that the thin string approximation (TSA) can be used, in which the string is treated as an infinitesimally thin line. The TSA enables semi-analytic estimates of equilibrium configurations~\cite{LEMPERIERE2003511}, but requires validation through full field-theoretic simulations, which long posed a significant computational challenge.

To alleviate this difficulty, kinky vortons were proposed~\cite{Battye2008KV} as $(2+1)$-dimensional analogues of vortons, which replace the string associated with the broken $U(1)$ symmetry with a kink (domain wall) in a $\mathbb{Z}_2 \times U(1)$ model. Kinky vortons were found to share the key qualitative features expected of their $(3+1)$-dimensional counterparts, while permitting treatment in a numerically tractable setting. Large radius configurations were well described by the TSA and the observed intervals of instability \cite{Battye2009SKV} were well captured by the elastic-string formalism of Carter and Martin \cite{CARTER1993151}. Kinky vortons, therefore, serve as a bridge between TSA predictions and full $(3+1)$-dimensional simulations, guiding subsequent confirmation of stable cosmic vorton solutions \cite{Battye:2021sji}, where it was shown that the TSA correctly predicted features of the solutions to a good degree of accuracy \cite{Battye:2021kbd}.

The two-Higgs-Doublet Model (2HDM) is a well motivated extension of the Standard Model (SM) which extends the scalar sector by adding a second Higgs doublet. This leads to five physical Higgs bosons: two CP-even, $h$ and $H$, one CP-odd, $A$, and a charged pair, $H^\pm$ \cite{PhysRevD_8_1226, Branco_2012}. This extended Higgs sector can accommodate additional sources of CP violation \cite{Keus_2016}, enable mechanisms for electroweak baryogenesis \cite{Fromme2006, Dorsch_2017}, and it also admits scenarios with an extra inert scalar that could serve as a dark matter candidate \cite{Grzadkowski_2009}. 

Beyond these phenomenological features, the 2HDM exhibits a rich topological structure \cite{Nishi:2006tg, Maniatis_2006, Ivanov:2006yq, Ivanov_2008, Battye2011VT, PILAFTSIS2012465}, as the model may exhibit a number of accidental global symmetries under different parameter restrictions. This allows for many varieties of topological defects to occur within the framework of the 2HDM, for example domain walls ($\mathbb{Z}_2$-symmetric)~\cite{Battye2011VT, ETO2018447, Battye2021SDW, PhysRevD.105.056007, BATTYE2025139311, CCDW}, strings ($U(1)$-symmetric)~\cite{Battye2011VT, Battye:2024iec, Battye:2024dvw}, and monopoles ($SO(3)_{\rm HF}$-symmetric)~\cite{Battye_2023, Battye:2024iec}. In this work we consider the $\mathbb{Z}_2$-symmetric case, and its simplification to the $U(1)$-symmetric case. Formation of these defects during the electroweak phase transition, due to the spontaneous breaking of the aforementioned symmetries, would produce cosmological relics which can act as a probes of high energy physics.

Numerical simulations of phase transitions in the global 2HDM \cite{Battye2021SDW, Battye_2023, Battye:2024iec, BATTYE2025139311, CCDW} have revealed that certain defects locally break the $U(1)_{\rm EM}$ symmetry at their cores, which would generate a non-zero photon mass in the corresponding gauged theory; throughout this work we refer to this simply as a non-zero photon mass. Corresponding one-dimensional solutions \cite{CCDW, Battye:2024iec, Battye_2023, Battye:2024dvw} confirm this behaviour for energy-minimizing field configurations. We use the standard terminology of current-carrying defects here, where ``superconducting'' refers to the existence of a localized charged condensate and persistent current, even in the absence of dynamical gauge fields. Recent work has shown that these defects can exhibit superconducting behaviour analogous to Witten’s $U(1) \times U(1)$ model, opening the door to vorton-like solutions in this beyond-the-Standard-Model setting. In the globally $U(1)$-symmetric case, stable superconducting strings carrying persistent bosonic currents have been demonstrated \cite{Battye:2024dvw}, establishing the necessary behaviour needed for vorton formation. Similarly, the $\mathbb{Z}_2$-symmetric variant admits current-carrying domain walls in $(2+1)$-dimensions \cite{CCDW}. In addition, a statistical study of the percolation of the $\mathbb{Z}_2$-symmetric vacuum \cite{BATTYE2025139311} identified long-lived ring-like configurations reminiscent of kinky vortons. These objects were seen to exist far longer than would be expected based on standard arguments of domain walls collapsing under their own tension.

Motivated by these developments, we investigate kinky vortons in the $\mathbb{Z}_2$-symmetric 2HDM as a computationally tractable proxy for 2HDM vortons. We present examples of stable ring solutions, demonstrate that the TSA accurately predicts their existence in the large-radius limit, and show that the elastic-string description correctly captures the radial dynamics of both stable and unstable configurations. Our analysis is restricted to the $\mathbb{Z}_2$-symmetric global theory, without SM gauge fields; nevertheless, the results provide a clear stepping-stone toward constructing fully gauged 2HDM vortons, which would constitute a novel heavy relic of the electroweak phase transition, with potentially observable consequences, in addition to possible mechanisms for baryogenesis.

\section{$\mathbb{Z}_2$-Symmetric 2HDM}\label{sec:theory}
For two complex Higgs doublets, $\Phi_1$ and $\Phi_2$, the general Lagrangian density of the global model takes the form
\begin{equation}
\mathcal{L} = |\partial^\mu \Phi_1|^2 + |\partial^\mu \Phi_2|^2 - V(\Phi_1,\, \Phi_2)
\end{equation}
where $|\partial^\mu \Phi_i|^2 = (\partial^\mu \Phi_i)^\dagger(\partial_\mu \Phi_i)$. There are multiple equivalent representations of both the fields and parameters of the 2HDM, some of which are more suitable for analytic calculations and others which provide simpler numerical implementation. An overview of the representations used in this work is provided in Appendix~\ref{sec:field_reps_and_params}.

The parameters of the model may be chosen such that the scalar potential, $V(\Phi_1,\, \Phi_2)$, can be symmetric under thirteen distinct accidental symmetries \cite{Battye2011VT, PILAFTSIS2012465}, giving rise to a rich topological structure of the vacuum. In this work we concern ourselves with the $\mathbb{Z}_2$-symmetric variant, which admits superconducting domain wall solutions in $(2+1)$-dimensions \cite{CCDW}, which are analogous to the superconducting string solutions found in the $U(1)$-symmetric variant \cite{Battye:2024dvw}. We note that the parameter restrictions defining the $U(1)$-variant correspond to a single additional constraint imposed on the $\mathbb{Z}_2$-symmetric potential.

Throughout this work, we always maintain the alignment limit of the theory, in which the properties of the SM Higgs boson are satisfied in agreement with experimental data \cite{Arbey_2018} and we choose the scalar particle $h$ to be identified as the SM Higgs boson. This condition is achieved by setting the mixing angles, $\alpha$ and $\beta$, (see Appendix~\ref{sec:field_reps_and_params}) to be equal; such that the two VEVs can be parametrised by $\tan\beta = v_2/v_1$. We additionally fix $M_h = 125$ GeV and $v_{\rm SM} = 246~\mathrm{GeV}$, where $v_{\rm SM} = \sqrt{v_1^2 + v_2^2}$ is the SM VEV \cite{Workman:2022ynf}.

Under these conditions, the $\mathbb{Z}_2$-symmetric scalar potential may be expressed in terms of the physical scalar masses as
\begin{align}
V =& -\half M_h^2\left(|\Phi_1|^2 + |\Phi_2|^2\right) \cr 
+& \frac{1}{2v_{\rm SM}^2}(M_h^2 + M_H^2\tan^2\beta)|\Phi_1|^4 \cr
+& \frac{1}{2v_{\rm SM}^2}(M_h^2 + M_H^2\cot^2\beta)|\Phi_2|^4 \cr
+& \frac{1}{v_{\rm SM}^2}\left[M_h^2 - M_H^2 + 2M_{H^\pm}^2\right]|\Phi_1|^2|\Phi_2|^2\cr
-& \frac{2}{v_{\rm SM}^2}M_{H^\pm}^2\Re(\Phi_1^\dagger \Phi_2)^2 \cr
+& \frac{2}{v_{\rm SM}^2}(M_A^2 - M_{H^\pm}^2)\Im(\Phi_1^\dagger \Phi_2)^2\,,
\label{eq:Z2_sym_pot}
\end{align}
while the $U(1)$-symmetric variant is an enhancement obtained by setting $M_A = 0$.

The topological defects of the model are most naturally characterised using the components, or combinations thereof, of the null 6-vector in the bi-linear field-space formalism \cite{Battye2011VT},
\begin{equation}
R^A  = \begin{pmatrix} {\Phi_1^\dagger \Phi_1 + \Phi_2^\dagger \Phi_2} \\ 
						{\Phi_1^\dagger \Phi_2 + \Phi_2^\dagger \Phi_1} \\ 
						{-i\left[\Phi_1^\dagger \Phi_2 - \Phi_2^\dagger \Phi_1\right]} \\ 
						{\Phi_1^\dagger \Phi_1 - \Phi_2^\dagger \Phi_2} \\
						\Phi_1^T i \sigma^2 \Phi_2 - \Phi_2^\dagger i \sigma^2 \Phi_1^* \\
						-i\left[\Phi_1^T i \sigma^2 \Phi_2 + \Phi_2^\dagger i \sigma^2 \Phi_1^*\right]
						\end{pmatrix}\,,
\label{eq:bi-linear_6_vec}
\end{equation}
where $\sigma^2$ is the second Pauli matrix. The topology of the $\mathbb{Z}_2$ vacuum admits domain walls in $R^1$, whereas the $U(1)$ vacuum admits string solutions in the combination $(R^1)^2 + (R^2)^2$. Both classes of defect generally exhibit additional structure, such as a localized violation of the so-called neutral vacuum condition, for which $R^\mu R_\mu \neq 0$ and the photon acquires a mass. The charged degrees of freedom are encoded in the complex scalar $\tilde R = R^4 + iR^5$, satisfying $R^\mu R_\mu = |\tilde R|^2$, while the scalar potential depends only on $R^\mu$, for $\mu = 0,1,2,3$.

Recent work has shown that all domain wall solutions of the $\mathbb{Z}_2$-symmetric theory can be grouped into distinct subclasses, each characterised by its own simplified ansatz \cite{CCDW}. Of relevance here is the subclass of superconducting solutions, which were found to admit superconducting walls in $(2+1)$-dimensions, with a current in $\tilde R$.

\section{Superconducting Domain Walls}\label{sec:super_con_sols}
Here we provide a brief summary of the superconducting domain walls identified in Ref.~\cite{CCDW}, which we use to construct kinky vortons in the following section. These solutions arise within a region of the 2HDM parameter space defined by
\begin{align}
	M_{H^\pm} < M_A\,, \quad M_{H^\pm}^2 \lesssim \half\left[M_H^2 +M_h^2\left(\frac{1}{f^{2}_1(0)} - 1\right) \right]\,,
\end{align}
where $f_1^2(0)$ may be evaluated numerically or semi-analytically (see Ref.~\cite{CCDW} for details). Within this region of parameter space the solutions are found to be independent of $M_A$, provided the first condition is met. The superconducting wall solutions form a good proxy for superconducting string solutions, in $(2+1)$-dimensions, as they effectively feel the same potential. The field configuration is fully described by the simplified ansatz
\begin{equation}
	\Phi = \frac{v_{\rm SM}}{\sqrt{2}}\begin{pmatrix}f_1 \sin\half\gamma_1  \\ f_1 \cos\half\gamma_1\\ f_+\cos\half\gamma_1 + f_2 \sin\half\gamma_1 \\ -f_+ \sin\half\gamma_1 + f_2 \cos\half\gamma_1\end{pmatrix} = \frac{v_{\rm SM}}{\sqrt{2}}\begin{pmatrix} g_1  \\ g_2\\ g_3 \\ g_4\end{pmatrix}\,,
	\label{eq:supercon_ansatz}
\end{equation}
where $\Phi = (\Phi_1\,, \Phi_2)^T$, and $f_+$ represents the superconducting condensate, for which a non-zero value signals neutral vacuum violation. The redefined fields $g_i$ provide both analytic and numerical simplifications in the work that follows.

To induce a current, this ansatz is acted upon by a space-time dependant transformation such that, for a $y$-directed domain wall,
\begin{equation}
	\Phi = \frac{v_{\rm SM}}{\sqrt{2}}\begin{pmatrix}g_1e^{i\left(\omega t + ky\right)}  \\ g_2\\ g_3e^{i\left(\omega t + ky\right)} \\ g_4\end{pmatrix}\,.
\label{eq:supercond_ansatz}
\end{equation}
This introduces an effective mass term into the Lagrangian, $-\half \kappa (g_1^2 + g_3^2)$, where $\kappa = \omega^2 - k^2$. A non-zero frequency, $\omega$, induces a charge, $Q=\omega L\int dx(g_1^2 + g_3^2)$, where $L$ is the length of the wall, and $k$, which describes the rate of winding of the condensate along the wall, generates a current.

\begin{figure*}
	\subfloat[]{\includegraphics[width=0.33\textwidth]{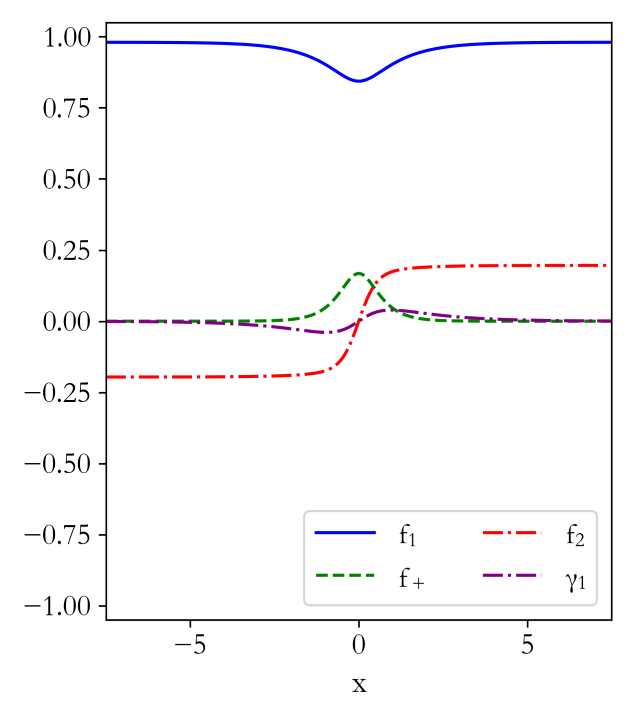}}
\subfloat[]{\includegraphics[width=0.33\textwidth]{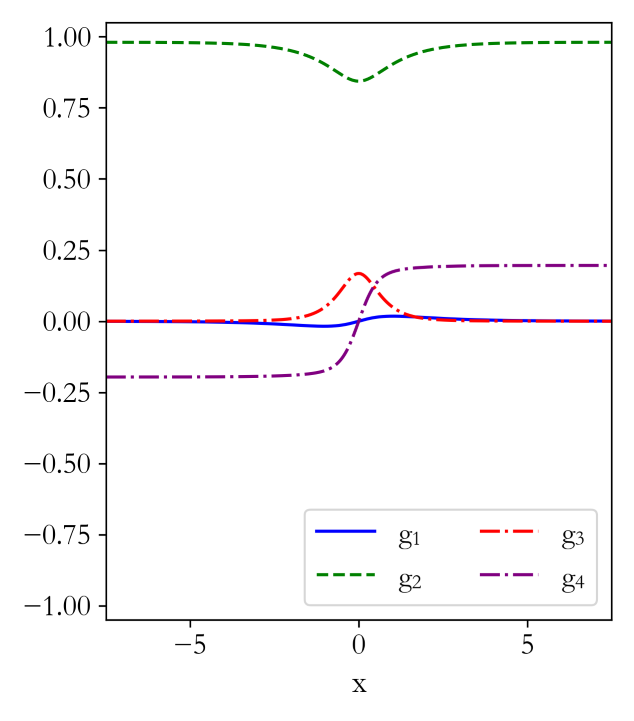}}
	\subfloat[]{\includegraphics[width=0.33\textwidth]{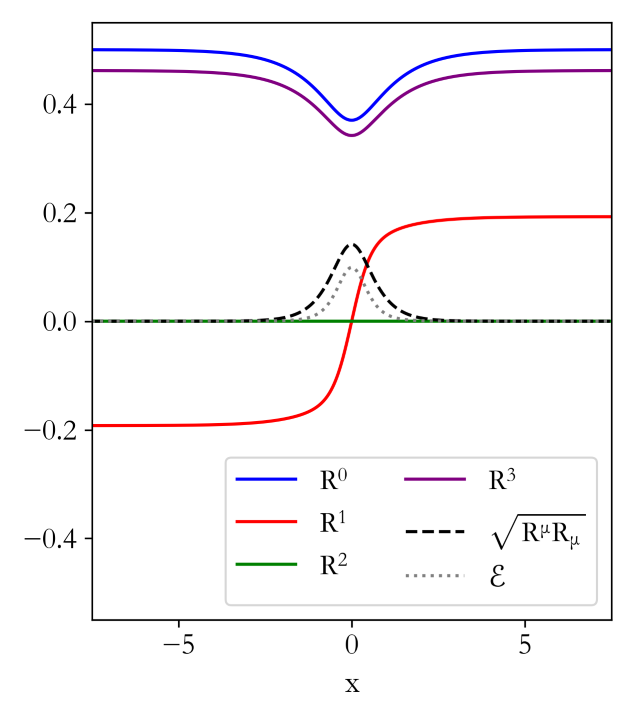}}
\caption{Example energy-minimizing kink solution for parameter set A in the $\mathbb{Z}_2$-symmetric global 2HDM, with $\kappa = -0.60$. Shown are the field profiles in the (a) general representation, (b) $g_i$ representation, and (c) bi-linear representation. Here, $R^4 = -\sqrt{R^\mu R_\mu}$ and hence $R^5 = 0$ globally. As $R^\mu R_\mu \neq 0$ at the centre of the kink, the $U(1)_{\rm EM}$ symmetry is broken locally.}
\label{fig:kink_sol_example}
\end{figure*}

Infinite walls constructed from static, one-dimensional kink solutions of this ansatz were shown to be stable to simple sinusoidal perturbations over long timescales, subject to stability criteria set by the longitudinal and transverse perturbation propagation speeds (see Sec.~\ref{sec:instab}). An example of such a kink solution is shown in Fig.~\ref{fig:kink_sol_example}, illustrating the behaviour of the field components.

These superconducting solutions were found to exist only in the magnetic regime ($\kappa <0$), as in the chiral ($\kappa=0$) and electric ($\kappa>0$) regimes the winding of the condensate would cause the vacuum, far from the defect, to contain non-zero gradient energies. We find no evidence to the contrary in the present study, and as such we concern ourselves only with the magnetic regime.

\section{Kinky Vortons}
We proceed by constructing kinky vortons in the 2HDM from the superconducting kink solutions described above. We present findings for four different parameter sets, chosen to illustrate qualitatively different kinky vorton dynamics; these are given in Table~\ref{tab:param_sets}. In particular, we find stable kinky vortons in parameter sets A and B, while parameter sets C and D admit configurations that are unstable to non-axially symmetric and longitudinal perturbations, respectively. Note that the specific parameter sets chosen are of no physical significance to the model, and are simply chosen to illustrate our points.

\begin{table}
    \centering
    \small
    \setlength{\tabcolsep}{3pt}
    \begin{tabular}{lccccc}
        \toprule
        Parameter Set & $M_H$ & $M_A$ & $M_{H^\pm}$ & $\tan\beta$ & $\kappa$ \\
        \midrule
	     A & 5000 & 1000 & 250 & 0.20 & -0.60 \\
            B & 875 & 500 & 125 & 0.25 & -0.10 \\
            C & 750 & 900 & 125 & 0.35 & -0.11 \\
            D & 600 & 300 & 200 & 0.85 & -0.12 \\
	\bottomrule
	\end{tabular}
    \caption{Masses in $\rm GeV$, and other parameters used in our analysis. Throughout we fix $M_h = 125\,{\rm GeV},\, v_{\rm SM}= 246\,{\rm GeV}$ and $\cos(\alpha - \beta) = 1$. Also given are the values of $\kappa$ for which we study kinky vorton solutions in each parameter set. Stable kinky vortons are found to exist in parameter sets A and B.}
    \label{tab:param_sets}
\end{table}

To construct kinky vortons within the 2HDM, we rely on two approximations: first, that they may be approximated as a segment of the infinite superconducting wall solution that has been wrapped into a loop, so that curvature effects are subdominant; and second, that the TSA holds, so that $\kappa$ does not vary significantly across the wall core. Under these assumptions, the equilibrium radius may be predicted from a straight superconducting wall solution. We will show, however, that these assumptions are not always satisfied in the $\mathbb{Z}_2$-symmetric global 2HDM, and their validity depends predominantly on $\kappa$ and the loop radius.

Following the semi-analytic method used in our previous work on superconducting walls \cite{CCDW}, based on that of Refs.~\cite{Battye:2021kbd,Battye2009SKV, CARTER1993151}, the total energy of a straight wall segment of length $L$ may be written as 
\begin{equation}
    E  =  \tau L  + \omega^2\Sigma_2L  =  \tau L + \frac{Q^2}{\Sigma_2L}\,,
    \label{eq:tot_wall_energy}
\end{equation}
where $\Sigma_2 = \int (g_1^2 + g_3^2) dx$ and, for this ansatz, $Q = \omega \Sigma_2 L$. The derivation of Eq.~\eqref{eq:tot_wall_energy} can found in Ref.~\cite{CCDW}, along with the definition of $\tau$ (see Eq. (45) of \cite{CCDW}), which represents the static energy per unit length of the wall. Differentiating the Lagrangian density of this ansatz with respect to the effective mass parameter $\kappa$ reveals that $\half\Sigma_2 = -\tau^\prime$, where $\prime$ represents differentiation with respect to $\kappa$. This allows us to construct the following quadratic,
\begin{align}
    & \left(\frac{L}{Q}\right)^4 + \frac{2}{\Sigma_2}\Bigg[\frac{\Sigma_2^\prime}{\Sigma_2^2} - \frac{1}{2\left(\kappa\Sigma_2 + \tau\right)}\Bigg] \left(\frac{L}{Q}\right)^2\nonumber \\
    & - \frac{2\Sigma_2^\prime}{\Sigma_2^4\left(\kappa\Sigma_2 + \tau\right)} = 0\,,
\end{align}
for which the relevant solution is
\begin{equation}
    \frac{L}{Q} = \sqrt{\frac{1}{\Sigma_2\left(\kappa\Sigma_2 + \tau\right)}}\,,
\end{equation}
while the remaining roots are found to be unphysical, see Ref.~\cite{Battye:2021kbd} for a detailed discussion. For a straight wall
\begin{equation}
    \kappa = \left(\frac{Q}{\Sigma_2 L}\right)^2 - \left(\frac{2\pi N}{L}\right)^2\,,
\end{equation}
which can be rearranged to show that
\begin{equation}
    \frac{L}{Q} = \sqrt{\frac{\Sigma_2^{-2} - \left(2\pi\frac{N}{Q}\right)^2}{\kappa}}\,,
\end{equation}
where $N$ is the integer winding number of the solution. Equating these two expressions for $L/Q$ reveals that a kinky vorton is expected to exist when the charge-to-winding ratio satisfies
\begin{equation}
    \frac{N}{Q} = \frac{1}{2\pi\Sigma_2}\sqrt{\frac{\tau}{\kappa\Sigma_2+\tau}}\,,
    \label{eq:N_Q_prediction}
\end{equation}
for which the equilibrium radius is predicted to be
\begin{equation}
R_* = N\sqrt{\frac{\Sigma_2}{\tau}}\,.
\label{eq:rad_prediction}
\end{equation}

\begin{figure}
    \centering
    \includegraphics[width=\columnwidth]{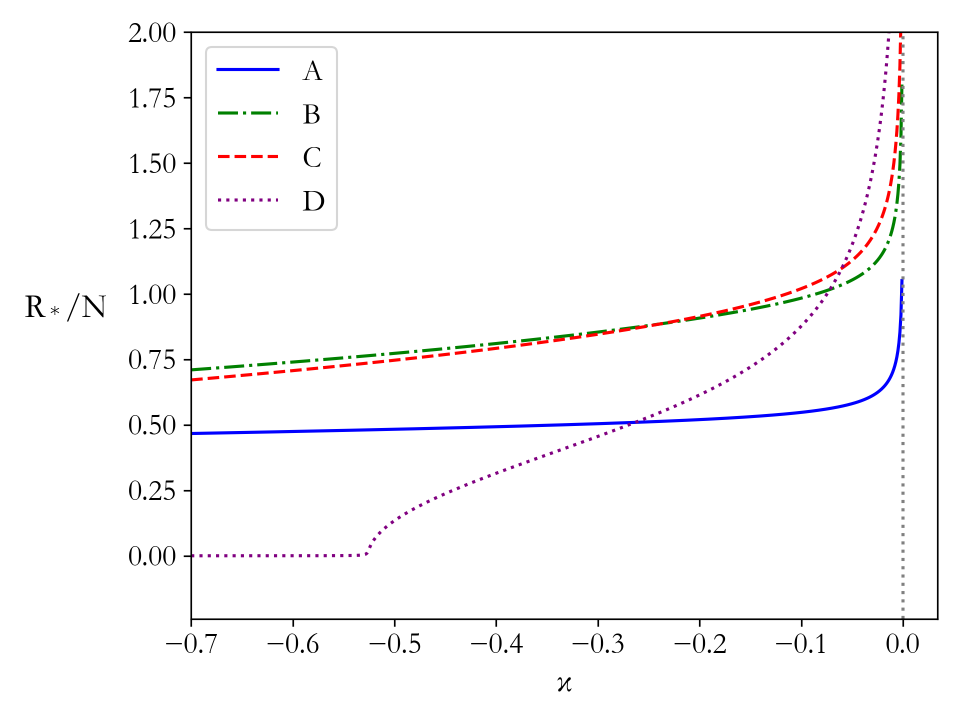}
    \caption{Predicted radius per winding number, $R_*/N$ as a function of $\kappa$ for $\kappa \in [-0.700, -0.001]$ for the parameter sets in Table~\ref{tab:param_sets}, using kink solutions computed with $n_x=4000,\, \Delta x=0.05,\, \delta=10^{-7}$ (see Appendix~\ref{sec:numericals-kinks} for definitions). Clear asymptotic behaviour is observed as $\kappa\to 0$.} 
    \label{fig:R_N_vs_kappa}
\end{figure}

We have evaluated $R_*/N$ for a range of $\kappa$ values in the magnetic regime, shown in Fig.~\ref{fig:R_N_vs_kappa}.
This predictive analysis further indicates that the ansatz is restricted to the magnetic regime, as $R_*/N$ diverges as $\kappa\to 0$ for all parameter sets we have considered.

\subsection{Instabilities}\label{sec:instab}
Before constructing kinky vortons, it is useful to identify the regions of $\kappa$-parameter space in which they are expected to be unstable.

\begin{figure}
    \centering
    \includegraphics[width=\columnwidth]{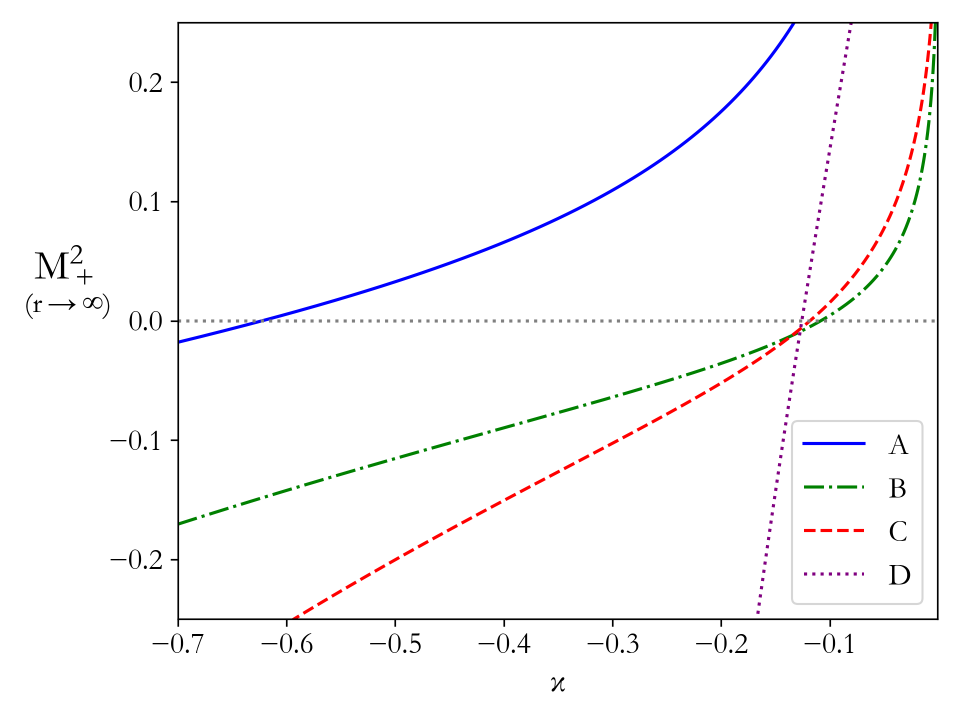}
    \caption{Effective mass of the condensate field, $f_+$, as $r \to \infty$, as a function of $\kappa$ for $\kappa\in[-0.700,-0.001]$ for the parameter sets in Table~\ref{tab:param_sets}, using kink solutions computed with $n_x=4000$, $\Delta x=0.05$, and $\delta=10^{-7}$ (see Appendix~\ref{sec:numericals-kinks} for definitions). Vacuum stability provides a clear lower bound on the $\kappa$-range in which kinky vortons may exist. For parameter set A, where the physical masses are larger than in sets B--D, this bound occurs at substantially larger $|\kappa|$. The corresponding upper bounds on $\omega$ are $\omega_A \le 1.961$, $\omega_B \le 0.970$, $\omega_C \le 0.944$, and $\omega_D \le 1.219$.}
    \label{fig:mass_stab}
\end{figure}

Firstly, we consider the stability of the vacuum state under the effective potential generated by the current and charge. This stability constraint was not considered in our work on superconducting walls, as $\kappa$ is spatially constant for a straight wall. For a vorton, however, $\kappa$ varies with radius according to $\kappa(r) = \omega^2 - (N/r)^2$. This means that as $r\to \infty$, the effective mass of the condensate field, $f_+$, becomes
\begin{equation}
\underset{(r \rightarrow \infty)}{M_+^2} = -\half \mu_2^2 + \half \lambda_2 f_2^2 + \quart \lambda_3 f_1^2 - \half \omega^2\,,
\label{eq:mass_instab}
\end{equation}
which must be positive for the vacuum to be stable and the charge to remain localized to the wall. An evaluation of $M_+^2(r\to\infty)$ is shown in Fig.~\ref{fig:mass_stab}, demonstrating that vacuum stability imposes a lower bound on $\kappa$ for all parameter sets considered.

Note that one could obtain an artificially stable kinky vorton where $M_+^2(r\to\infty) <0$ by placing the boundaries of the simulation sufficiently close to the defect; however, such a solution would be unphysical and any large enough simulation would reveal the instability \cite{Battye:2021kbd}.

\begin{figure}
    \includegraphics[width=\columnwidth]{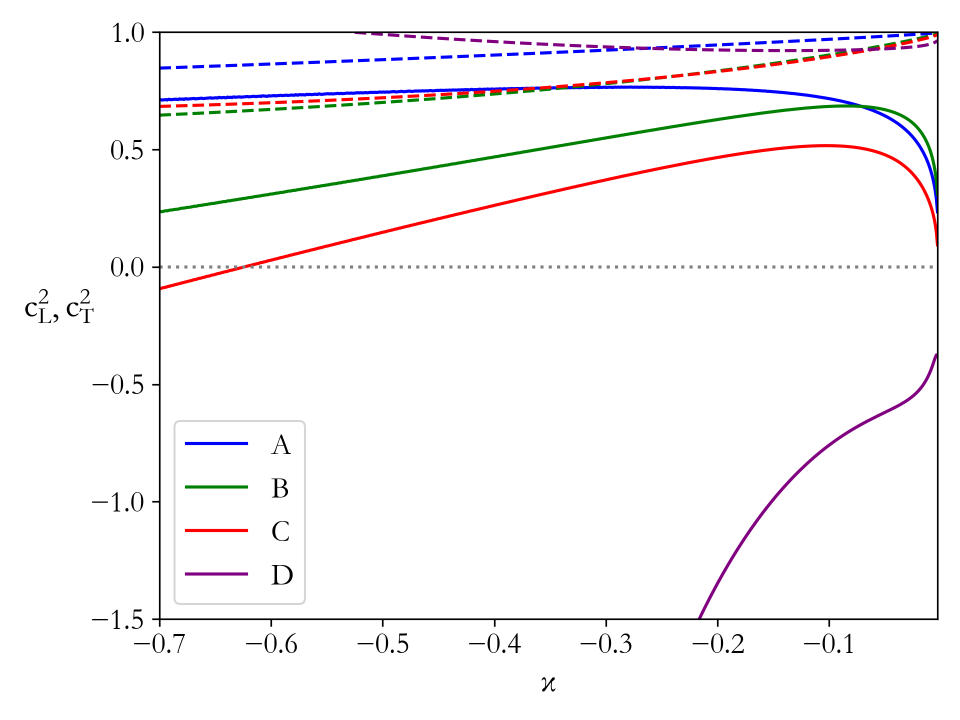}
    \caption{Longitudinal and transverse perturbation propagation speeds, $c_L^2$ (solid) and $c_T^2$ (dashed), for current-carrying walls in the 2HDM as a function of $\kappa$ for $\kappa\in[-0.700,-0.001]$ for the parameter sets in Table~\ref{tab:param_sets}. Kink solutions were computed with $n_x=4000$, $\Delta x=0.05$, and $\delta=10^{-7}$ (see Appendix~\ref{sec:numericals-kinks} for definitions). A second-order finite-difference scheme is used to evaluate $\Sigma_2^\prime$ for the calculations of $c_L^2$ and $c_T^2$. We see non-standard behaviour in $c_L^2$ as $\kappa \to 0$, as Eq.~\eqref{eq:speeds} would naively suggest that $c_L^2 \to 1$.} 
    \label{fig:prop_speeds}
\end{figure}

We now consider the stability of the defects themselves. Following Ref.~\cite{CCDW}, we first examine the longitudinal and transverse perturbation propagation speeds, $c_L^2$ and $c_T^2$. These quantitates can be found by diagonalizing the energy-momentum tensor to find the equation of state, see Refs.~\cite{Battye:2021kbd, Battye2009SKV} for further details. In the magnetic regime of our ansatz they are given by
\begin{equation}
    c_L^2 = 1 + \frac{2\kappa\Sigma_2^\prime}{\Sigma_2}\,, \quad c_T^2 = 1+\frac{\kappa\Sigma_2}{\tau}\,.
\label{eq:speeds}
\end{equation}
For a superconducting wall, whether infinite or closed into a loop, stability and causality require $0<c_L^2<1$ and $0<c_T^2<1$. These speeds are shown in Fig.~\ref{fig:prop_speeds}, where we can already see there will be no stable kinky vortons in parameter set D, as $c_L^2 < 0$ for all $\kappa$. We observe a non-standard behaviour for current-carrying defects here, in that as $\kappa \to 0$ $c_L^2 \not\to 1$. This provides further evidence that this ansatz is restricted to the magnetic regime in the global 2HDM: as $\kappa\to 0$ one finds $\Sigma_2\to\infty$ because $\gamma_1$ does not approach zero in the vacuum in the non-magnetic regimes.
\FloatBarrier

\begin{figure*}
    \centering
    \subfloat[Parameter set A]{\includegraphics[width=0.5\textwidth]{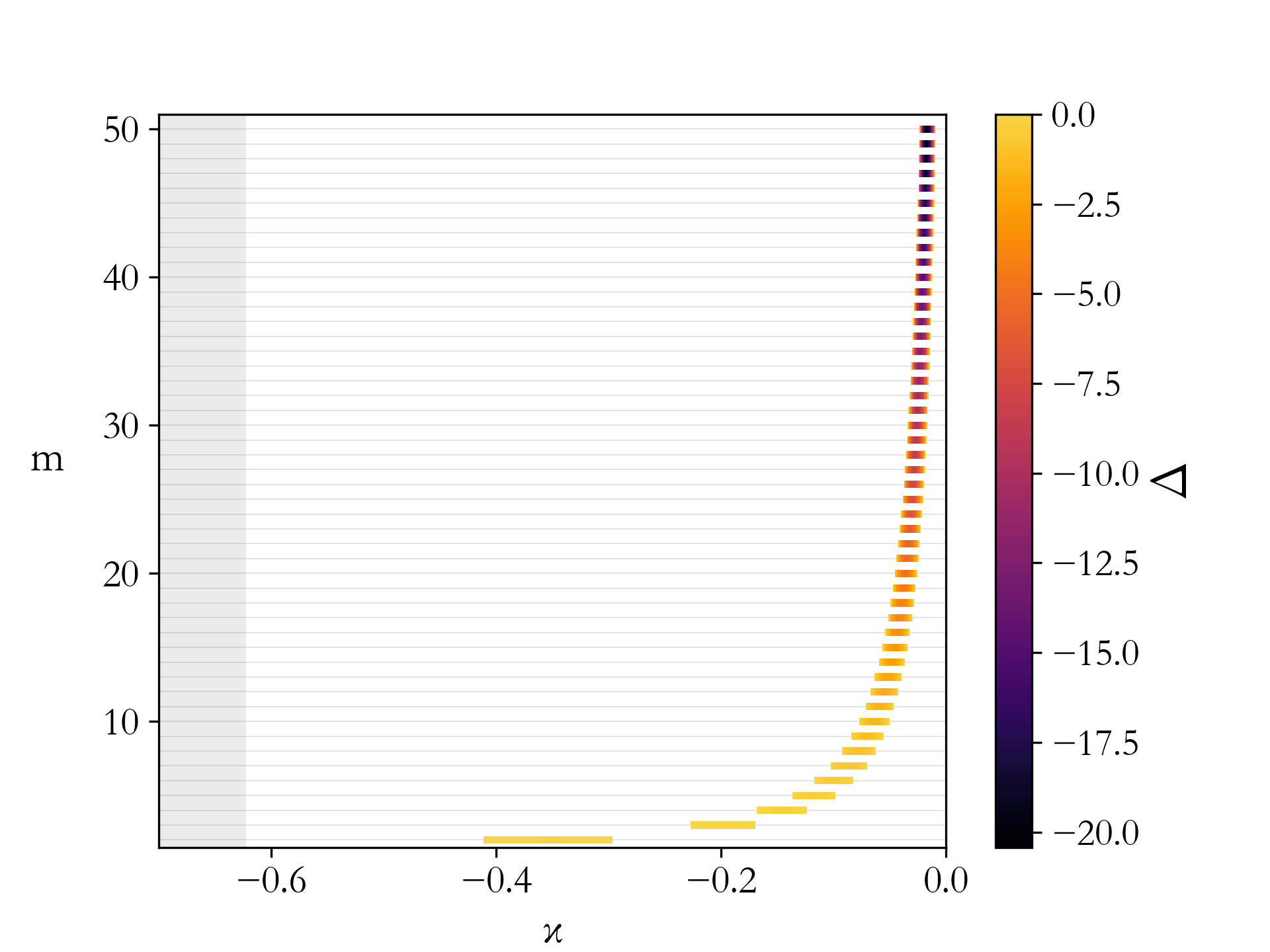}}
    \subfloat[Parameter set B]{\includegraphics[width=0.5\textwidth]{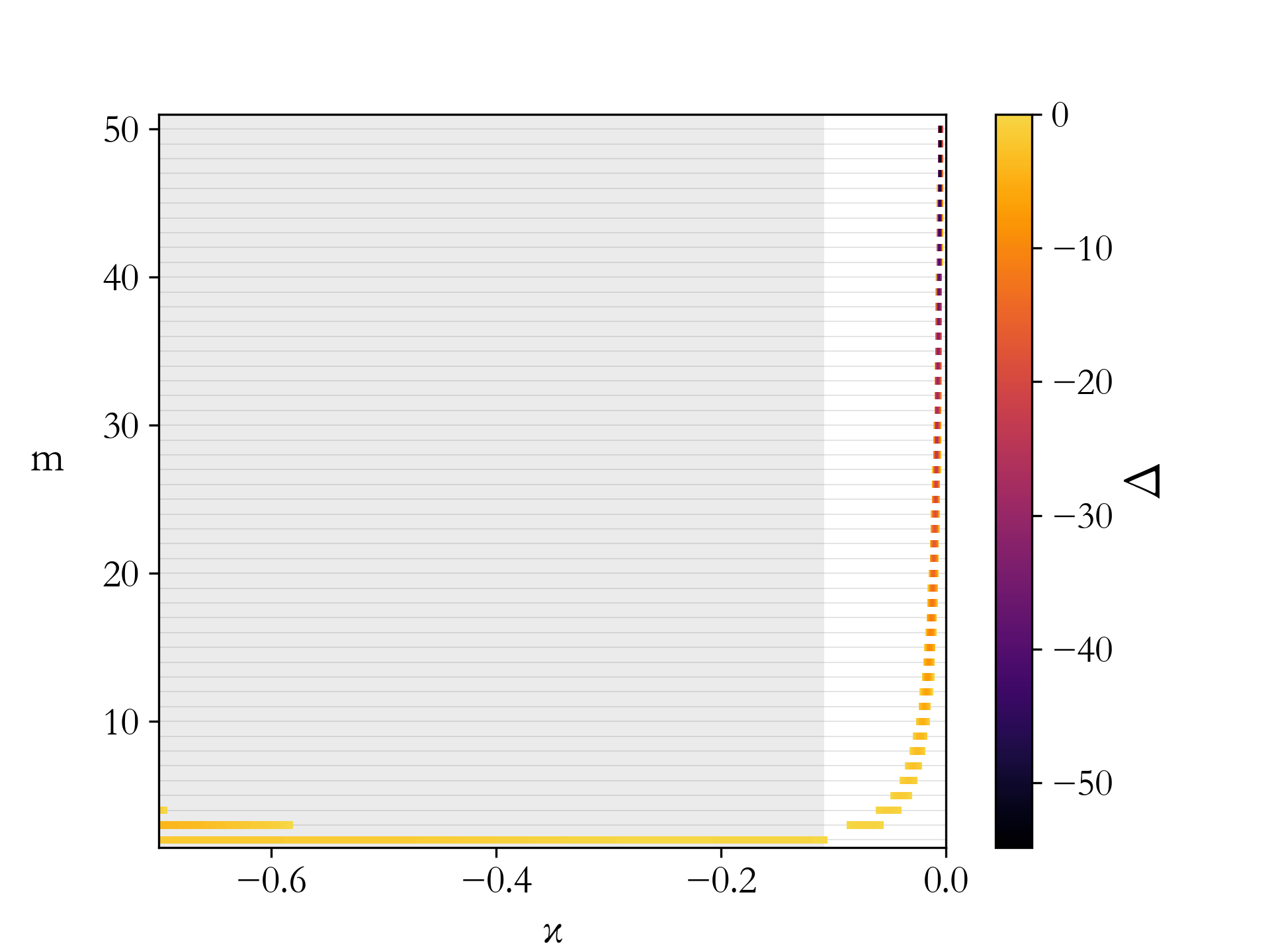}}
    \\\vspace{-10pt}
    \subfloat[Parameter set C]{\includegraphics[width=0.5\textwidth]{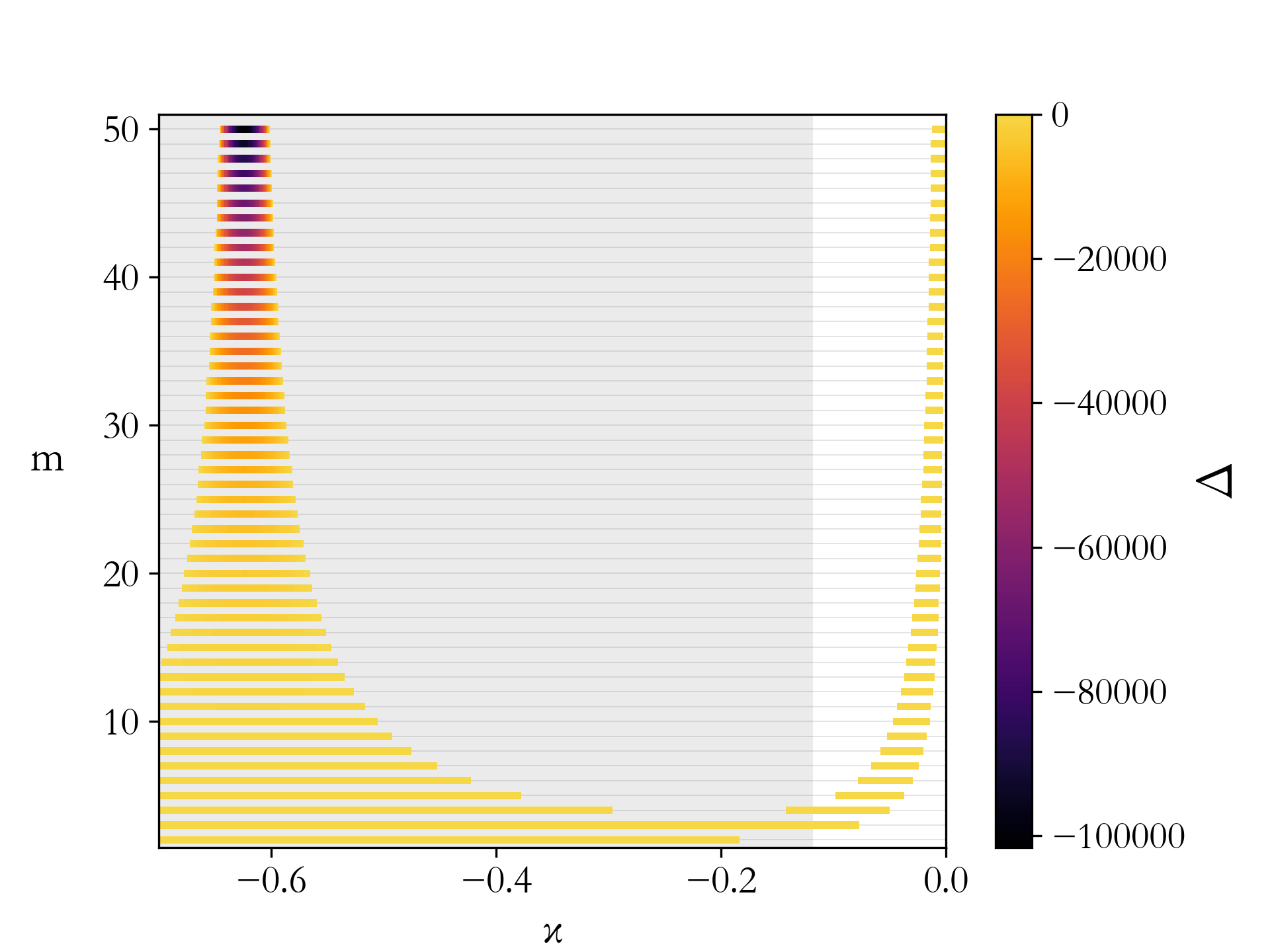}}
    \subfloat[Parameter set D]{\includegraphics[width=0.5\textwidth]{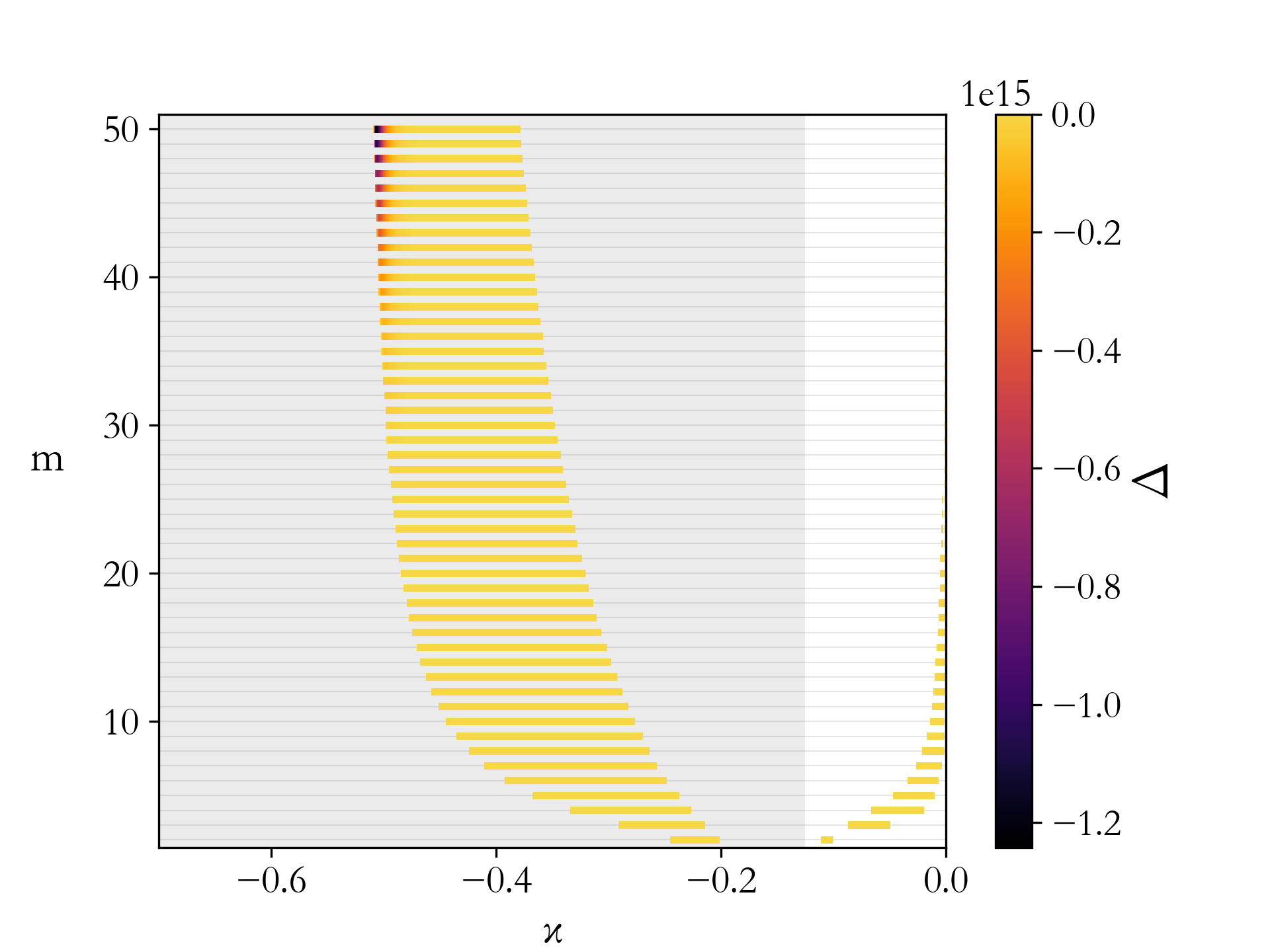}}
    \caption{Intervals of instability for $\kappa\in[-0.700,-0.001]$ for different modes of oscillation ranging from $m=2$ to $m=50$. The coloured regions show where the discriminant of the cubic is less than zero, and as such values of $\kappa$ for which kinky vortons would be expected to be unstable to a given vibrational mode. Shaded areas of the plots correspond to regions of vacuum instability. There exist regions of stability in parameter sets A and B, when their vacuum instability is also considered, parameter set C is always unstable and, despite appearances, parameter set D does not have regions of stability because $c_L^2$ is always negative, see Fig.~\ref{fig:prop_speeds}.}
    \label{fig:mode_eval}
\end{figure*}

The propagation speeds can be used to determine stability with respect to vibrational modes of order $m$. Perturbations of a circular loop of superconducting wall (or string) of radius $R$, stabilised about an equilibrium configuration by the centrifugal effect of the current, may be expressed in terms of one transverse variable $\tilde\beta$ and two longitudinal variables $\tilde\alpha$ and $\tilde\epsilon$ \cite{CARTER1993151}. Specifically, these variables satisfy the eigenvalue equation
\begin{widetext}
\begin{equation}
{\setlength{\arraycolsep}{10pt}
\begin{pmatrix}
2 & c_T^2 + c_L^2 & (1+c_L^2)\nu_m - 2m \\
(1+c_T^2)\nu_m - 2m & c_T^2(c_L^2 +1)\nu_m - (c_T^2 + c_L^2)m & 2 \\
(1-c_T^2)\nu_m & c_T^2(c_L^2 -1)\nu_m - (c_T^2 - c_L^2)m & 0 
\end{pmatrix}}
\begin{pmatrix}
c_T\tilde\epsilon \\
\tilde \alpha \\
i\tilde \beta R
\end{pmatrix} = 0\,,
\label{eq:osc_eigen}
\end{equation}
\end{widetext}
where $\nu_m = \Omega_m R/c_T$, and $\Omega_m$ are the perturbation frequencies (see refs.~\cite{Battye:2021kbd, Battye2009SKV, CARTER1993151} for further  details). The determinant of such an eigenvalue equation must vanish, and hence
\begin{equation}
    a_3\nu_m^3 + a_2\nu_m^2 + a_1\nu_m + a_0 = 0\,,
\label{eq:freq_cubic}
\end{equation}
with
\begin{align}
a_0 &= 2(c_L^2 - c_T^2)(m^2 - 1)m\,, \nonumber \\
a_1 &= 4c_T^2(1 - c_L^2)(m^2 - 1) - (1 + c_T^2)(c_L^2 - c_T^2)(m^2 + 1)\,, \nonumber \\
a_2 &= 2c_T^2[c_L^2 - c_T^2 - 2(1 - c_L^2 c_T^2)]m\,, \nonumber \\
a_3 &= c_T^2(1 + c_T^2)(1 - c_L^2 c_T^2)\,.
\end{align}
Stable oscillations for a given mode require real roots. Therefore, if the discriminant,
\begin{equation}
    \Delta = a_1^2a_2^2 - 4a_1^3a_3 - 4a_0a_2^3 - 27a_0^2a_3^2 + 18a_0a_1a_2a_3\,,
\end{equation}
is negative, the vorton will be unstable to the corresponding mode. Axially symmetric oscillations and translations correspond to $m=0, \,1$ respectively, and are stable provided $0<c_L^2<1$ and $0<c_T^2<1$. For modes with $m\ge 2$, stability is determined by the sign of $\Delta$. Having computed the propagation speeds for each parameter set, we use this criterion to identify the intervals of instability as a function of $\kappa$, which we present in Fig.~\ref{fig:mode_eval} for each of the parameter sets A--D.

Combining the perturbation-stability intervals with the vacuum-stability constraint from $M_+^2$, we predict that parameter set A admits stable kinky vortons for $\kappa\in[-0.622,-0.412]$, $\kappa\in[-0.296,-0.228]$, and a very narrow window around $\kappa = -0.169$, while parameter set B admits stable solutions for $\kappa\in[-0.105,-0.089]$. By contrast, parameter set C is predicted to be unstable to non-axially symmetric perturbations throughout the vacuum-stable region, and parameter set D is likewise predicted to be unstable, but to longitudinal perturbations, which can be seen from Fig.~\ref{fig:prop_speeds}.

In agreement with the vortons of Ref.~\cite{Battye:2021kbd}, we find that regions of absolute stability correspond to those where the longitudinal speed is not significantly less than the transverse speed, as seen when comparing Fig.~\ref{fig:prop_speeds} with the regions of stability for parameter sets A and B. We emphasise, however, that these stability predictions rely on the TSA holding, which is expected for sufficiently large kinky vortons. In our ansatz the effective wall width is predominantly set by the profile of $\gamma_1$, which depends on $|\kappa|$; consequently, small-radii vortons are expected to be comparatively rare in the parameter space of the model.

\subsection{Vorton solutions and dynamics}

For a circular current--carrying loop, we adopt the polar--coordinate ansatz
\begin{equation}
\Phi  =  \frac{v_{\rm SM}}{\sqrt{2}}\begin{pmatrix}g_1e^{i\left(\omega t + N\theta\right)}  \\ g_2\\ g_3e^{i\left(\omega t + N\theta\right)} \\ g_4\end{pmatrix}\,,
\label{eq:CC_DW}
\end{equation}
where the radial profiles $g_i \equiv g_i(r)$ are taken from the corresponding kink-solution of  Eq.~\eqref{eq:supercond_ansatz}, with the identification $g_i(r=R)=g_i(x=0)$ for a loop of radius $R$.

\subsubsection{Solutions}

For a chosen winding number $N$ and $\kappa$, the energy minimising charge $Q$ is determined by Eq.~\eqref{eq:N_Q_prediction} and the predicted equilibrium radius $R_*$ follows from Eq.~\eqref{eq:rad_prediction}. To construct an initial kinky vorton configuration we apply a period of gradient flow to the TSA-predicted profile, see Appendix~\ref{sec:numericals-GF} for further details. This allows the solution to approach the energy minimizing loop configuration, which will never be perfectly described by the straight wall profile.

The initial frequency is taken to be $\omega_i = \sqrt{\kappa + (N/R_*)^2}$, and is allowed to vary during gradient flow to conserve the total charge $Q$. For those values of $N$ for which gradient flow converges, the resulting configuration subsequently evolves as a stationary kinky vorton under full $(2+1)$-dimensional dynamics, provided no dynamical instabilities are present. Below a critical value of $N$, gradient flow fails to converge at the predicted winding-to-charge ratio as the TSA begins to break down.\footnote{Additionally, we performed systematic scans of the charge $Q$ around the thin-string prediction for $N/Q$, at a given $N$. In all cases, the optimal value of $Q$ was found to be practically that of the thin-string estimate, with only marginal improvements in the residuals of the equations of motion for minimal variations in $Q$. When gradient flow failed to converge at the thin-string prediction, no nearby solutions were obtained by varying $Q$, reflecting the breakdown of the TSA below a critical $N$.} While this does not preclude the existence of kinky vortons of smaller $N$ and $R$, whose structure is not well described by the TSA, finding such solutions would require a substantially broader exploration of initial conditions. Since the present study is intended as a proof of concept within the $\mathbb{Z}_2$-symmetric global 2HDM, we choose to only focus on those solutions well described by the semi-analytic method we have presented.

\begin{figure*}
	\subfloat[$N=1000$, $Q=174.5$]{\includegraphics[width=0.33\textwidth]{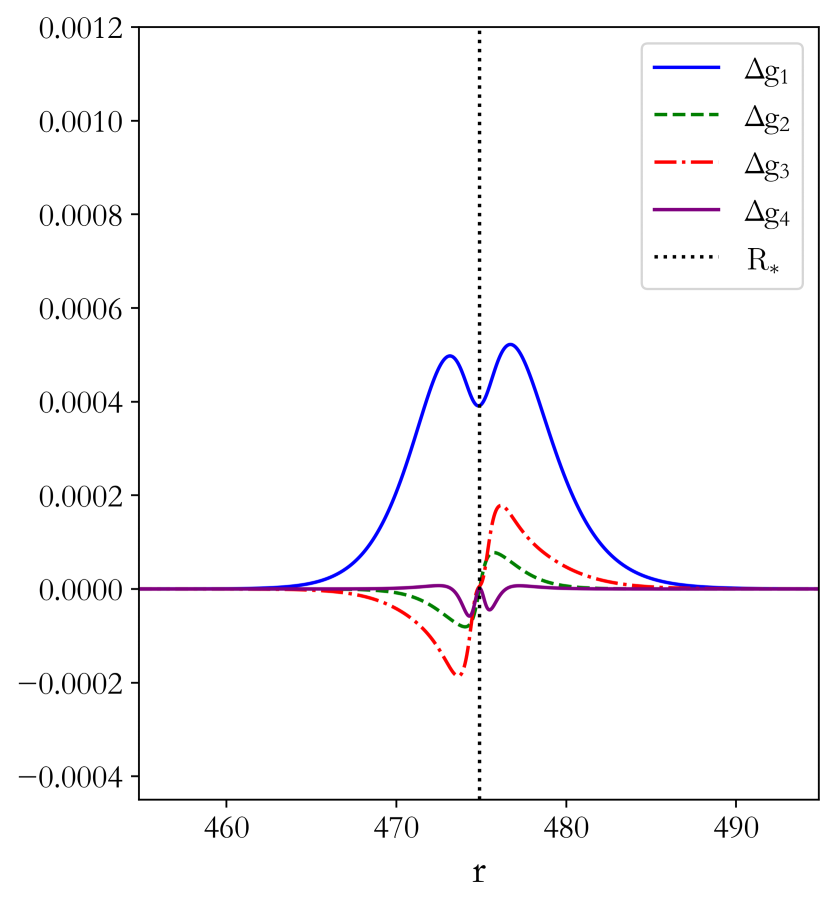}}
\subfloat[$N=750$, $Q=130.9$]{\includegraphics[width=0.33\textwidth]{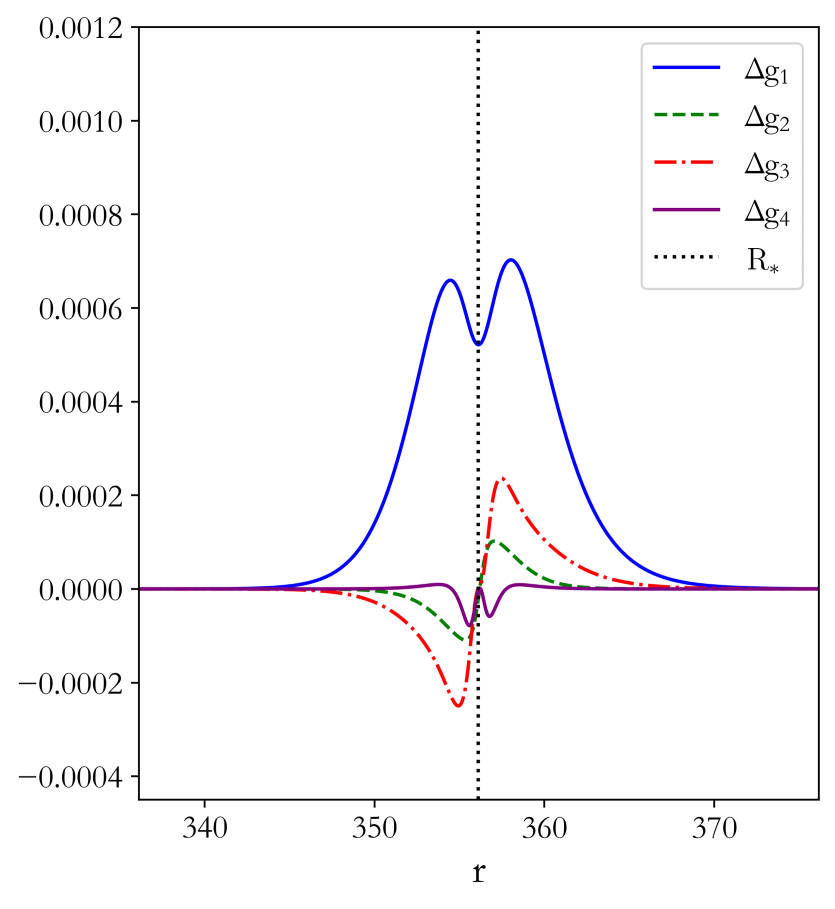}}
	\subfloat[$N=500$, $Q=87.2$]{\includegraphics[width=0.33\textwidth]{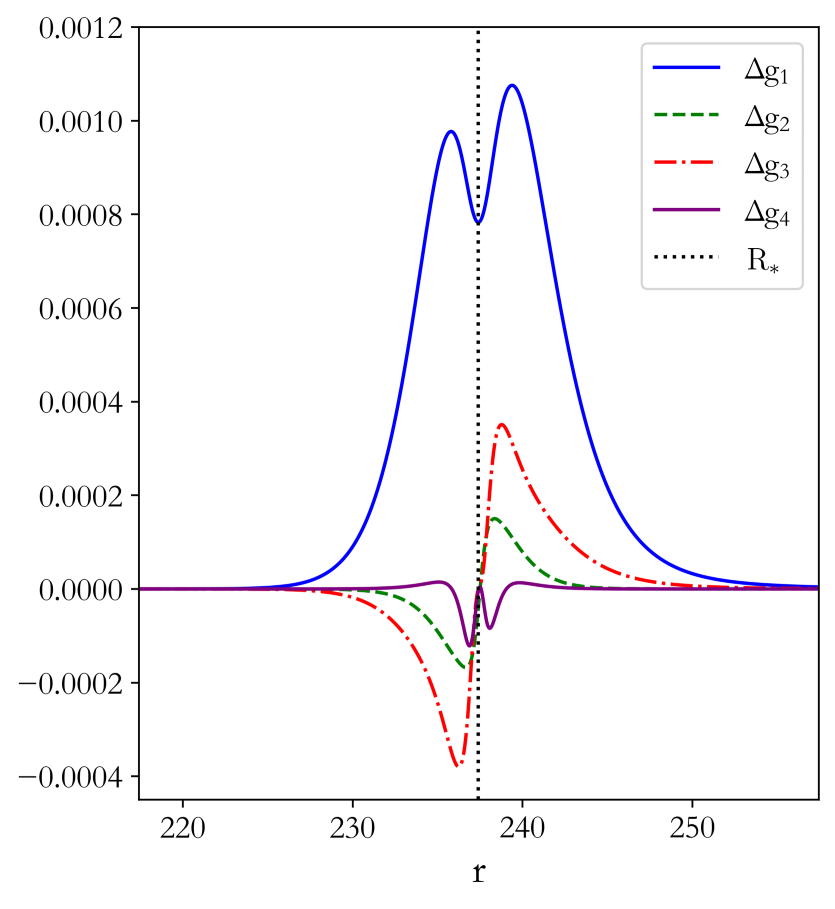}}
\caption{Comparison between relaxed radial-profile kinky vorton solutions in parameter set A and the profiles obtained by placing the straight-wall TSA prediction at $R_*$. As $N$ decreases, the relaxed solutions develop increasing width and asymmetry. Also shown are the predicted radii of the solutions, $R_*$.}
\label{fig:grad_flow_diff}
\end{figure*}

We find that the termination point of the kinky vorton solution branches are strongly parameter-set dependent. For example, parameter set A admits kinky vorton solutions down to approximately $N=500$ with radius $R \simeq 237$, whereas for parameter set B the branch terminates at approximately $N=800$ and $R \simeq 788$. This behaviour is consistent with the parameter dependence of the intrinsic width of the superconducting wall solutions, in particular the spatial variation of $\gamma_1$ (see Ref.~\cite{CCDW} for further details), which depends on both the parameter set and value of $\kappa$.

Figure~\ref{fig:grad_flow_diff} compares the relaxed radial profiles obtained via gradient flow with those obtained by simply placing the kink solution at the TSA-predicted radius $R_*$. Across all the parameter sets studied, the relaxed kinky vorton profiles become wider and increasingly asymmetric as $N$ decreases. This reflects the onset of a splitting instability should $N$ and subsequently $Q$ not be sufficiently large, as described in Ref.~\cite{Battye2008KV}, where the effective potential trapping the condensate on the loop becomes too weak to prevent loss of charge and the condensate radiates away to infinity. In the present model, the critical values of $N$ (and hence $R$) are substantially larger than those found in Ref.~\cite{Battye2008KV}, however this could well reflect the strict parameter dependence of these solutions. It will, therefore, be important to revisit this question in the gauged theory, where we expect both the magnetic-regime-restriction and the long-range gradient energy associated with the phase winding to be modified.

\subsubsection{Stable vortons}

We use the relaxed radial profiles to construct kinky vortons by mapping them to the fields of the linear representation at a given distance from the centre of the loop, with the phase and current initially imposed through the $(t,\theta)$ dependence in Eq.~\eqref{eq:CC_DW}, using the frequency of the relaxed radial profile. We then evolve these configurations under $(2+1)$-dimensional dynamics using the numerical methodology described in Appendix~\ref{sec:numericals-sims}.

\begin{figure*}
\includegraphics[width=\textwidth]{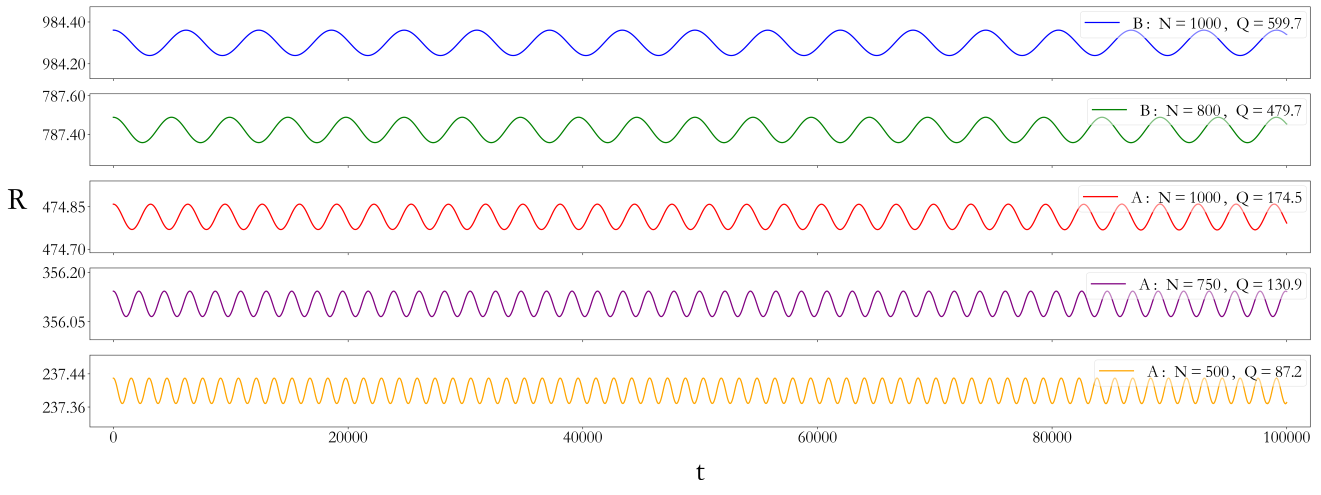}
\caption{Evolution of the core of a domain wall at fixed $\theta=0$ during $(2+1)$-dimensional dynamics for kinky vortons of varying winding number $N$ and charge $Q$ initialised in circularly symmetric states. All solutions exhibit bounded radial oscillations with frequencies consistent with the elastic-string $m=0$ breathing mode, with normalised oscillation amplitudes below $0.1\%$ of the average and no evidence of growing modes. A radial lattice size of $N_r = 1000$ is used in all above simulations, with $N_\theta = 30976,\, 24768,\, 14976\,, 11200\,, 7488$ (top to bottom), such that $\Delta r \approx r\Delta\theta$ in the vicinity of the defect.}
\label{fig:stab_vort_long}
\end{figure*}

\begin{figure}
\centering
\subfloat[$t=0$]{\includegraphics[width=0.439\columnwidth]{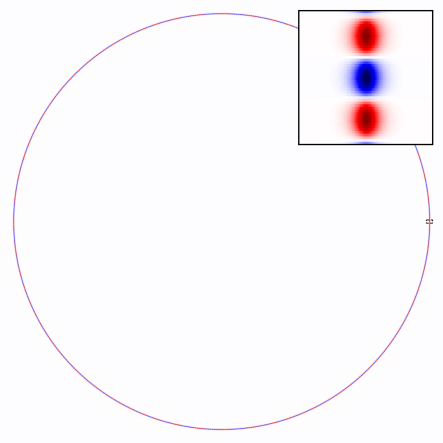}}\hspace{1pt}
\subfloat[$t=100000$]{\includegraphics[width=0.439\columnwidth]{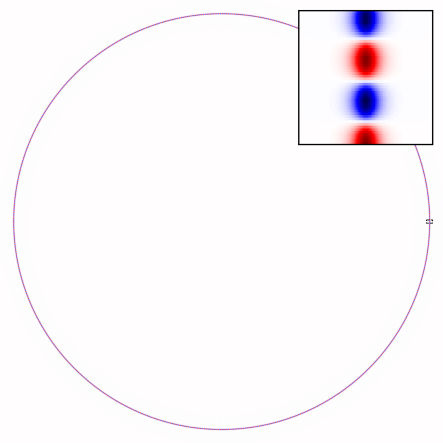}}\hspace{1pt}
\includegraphics[width=0.09\columnwidth]{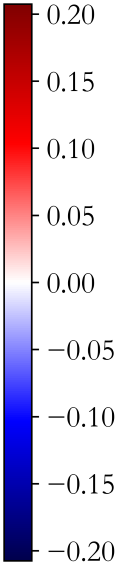}
\caption{Snapshots of the bi-linear component $R^4$, which contains the charged degrees of freedom, from the evolution of a kinky vorton in parameter set A with $\kappa=-0.60$, $N=500$, $Q=87.2$, $R=237.4$, and $\omega=1.958$. The condensate winding is not resolved in the full-frame images; the inset panels show close-ups of the wall centre at $\theta=0$, illustrating the persistent current that flows around the loop. Lattice sizes are $N_r = 1000$ and $N_\theta = 7488$.}
\label{fig:param_A_long_evo}
\end{figure}

In Sec.~\ref{sec:instab} we predicted stable kinky vortons to exist in parameter sets A and B. Figure~\ref{fig:stab_vort_long} shows the radial evolution of several kinky vortons with varying $N$ and $Q$ in both parameter sets, initialised in circularly symmetric states. The $\kappa$ values have been chosen such that the elastic string description predicts stability to all vibrational modes.
All kinky vortons shown present completely bounded radial oscillations, of frequencies consistent with the predicted $m=0$ ``breathing'' modes of the elastic string description, which we summarize in Table~\ref{tab:stab_vort_freq}. We see that the TSA predicts the radii of these kinky vortons with high accuracy, and that the elastic string description perfectly captures the natural oscillation frequency of unperturbed solutions.
\begin{table}
    \centering
    \small
    \setlength{\tabcolsep}{3pt}
    \begin{tabular}{lcccccc}
        \toprule
        Parameter Set & $N$ & $Q$ & $R$ & $\kappa$ & $f/10^{-4}$ & $f_0/10^{-4}$\\
        \midrule
	     B & 1000 & 599.7 & 984.4 & -0.1 & 1.60 & 1.60 \\
	     B & 800 & 479.7 & 787.5 & -0.1 & 2.02 & 2.01 \\
	     A & 1000 & 174.5 & 474.9 & -0.6 & 3.31 & 3.30 \\
	     A & 750 & 130.9 & 356.1 & -0.6 & 4.40 & 4.41 \\
	     A & 500 & 87.2 & 237.4 & -0.6 & 6.60 & 6.61 \\
	\bottomrule
	\end{tabular}
    \caption{Representative stable kinky vorton solutions in the 2HDM, showing measured oscillation frequencies $f$ from unperturbed states and the corresponding elastic-string breathing-mode predictions $f_0$. Note that $R$ is found to be consistent with the TSA predicted $R_*$ -- the fractional error is $\sim 10^{-6}$.}
    \label{tab:stab_vort_freq}
\end{table}
Figure~\ref{fig:param_A_long_evo} shows snapshots of the evolution of the smallest kinky vorton with $N=500$ found in parameter set A, as an example.

\begin{figure*}
\centering
	\subfloat[$m=2$]{\includegraphics[width=0.33\textwidth]{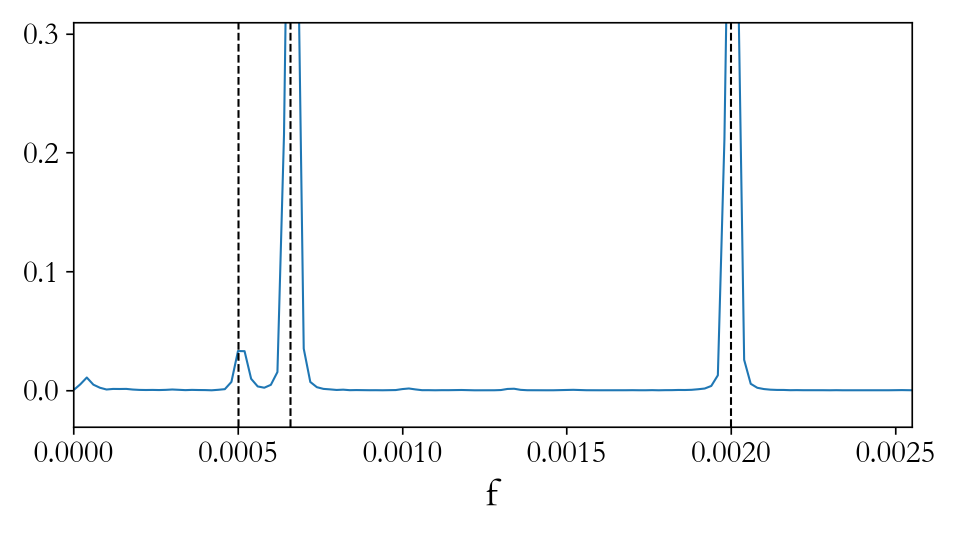}}
	\subfloat[$m=3$]{\includegraphics[width=0.33\textwidth]{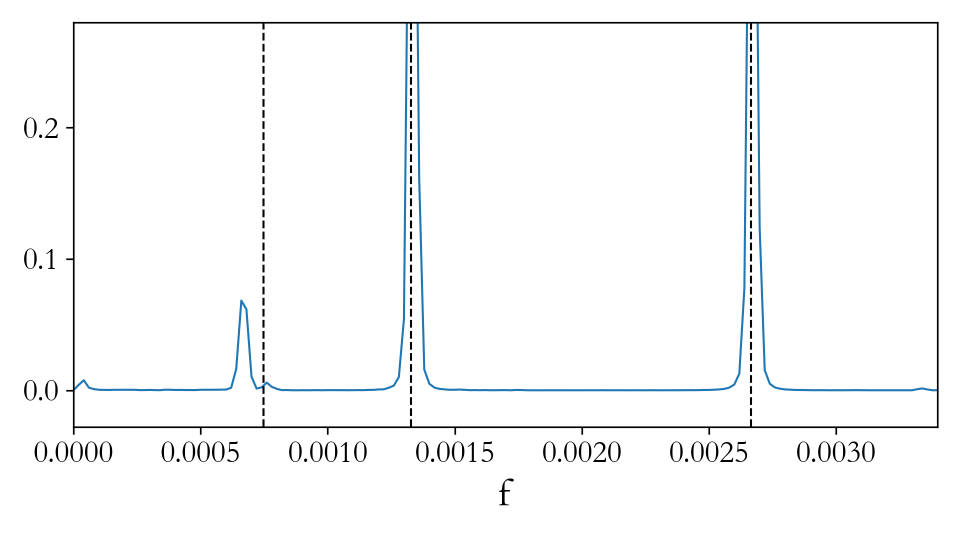}}
	\subfloat[$m=4$]{\includegraphics[width=0.33\textwidth]{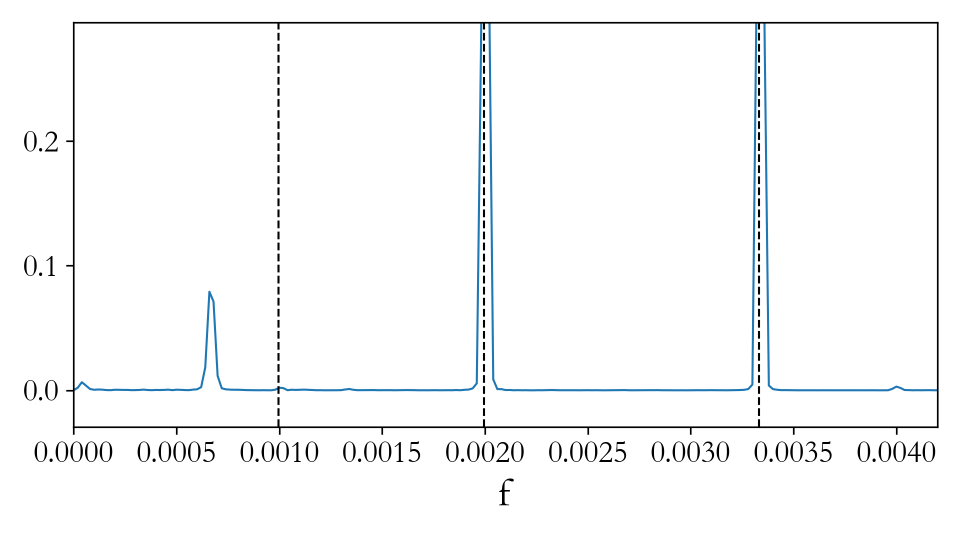}}\\
	\subfloat[$m=5$]{\includegraphics[width=0.33\textwidth]{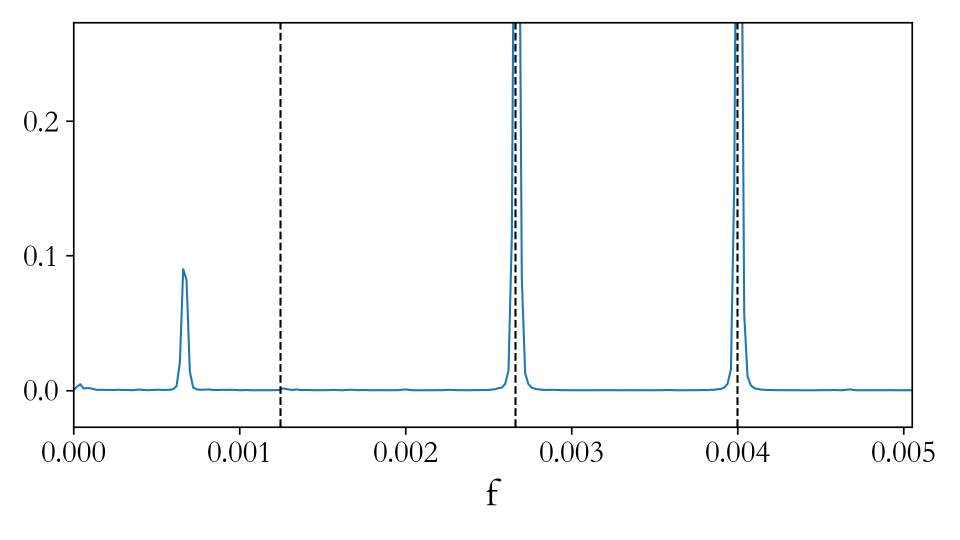}}
	\subfloat[$m=6$]{\includegraphics[width=0.33\textwidth]{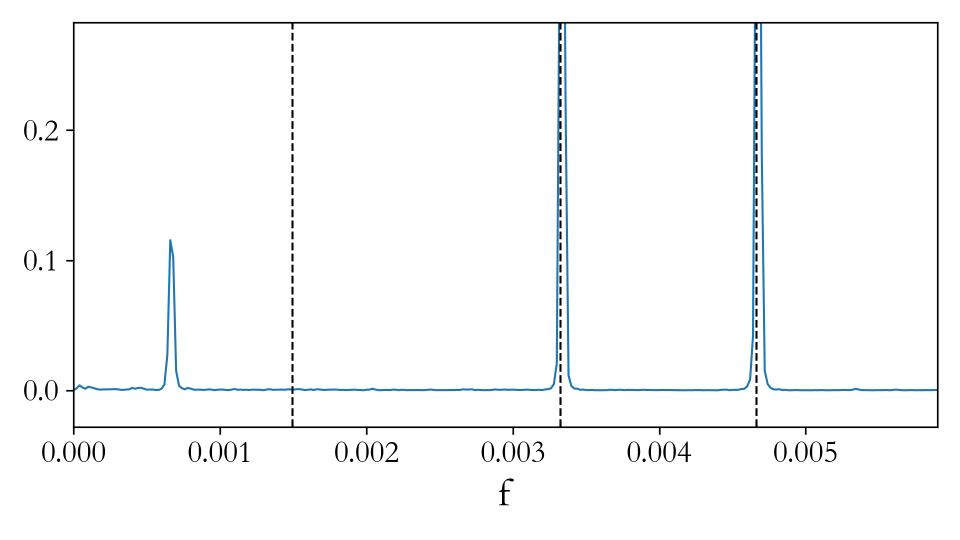}}
	\subfloat[$m=7$]{\includegraphics[width=0.33\textwidth]{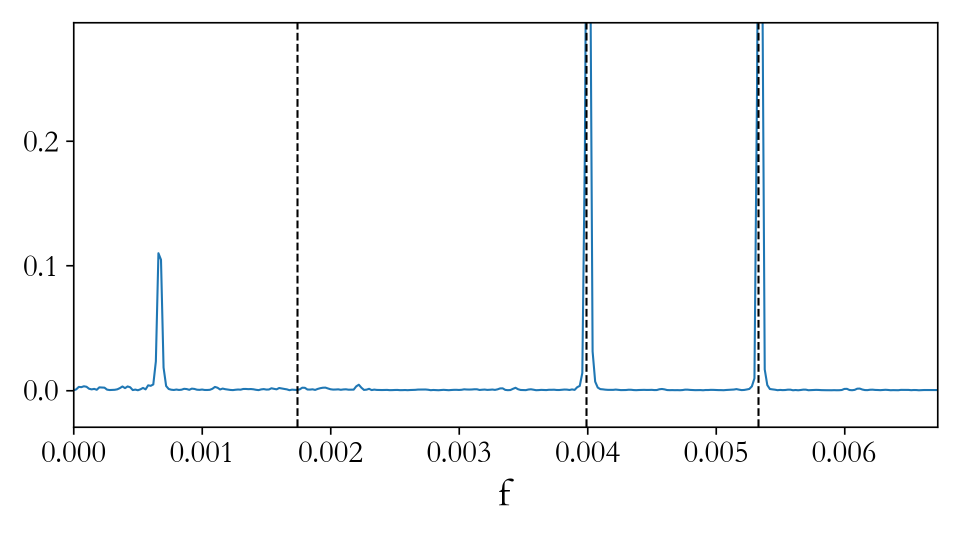}}\\
	\subfloat[$m=8$]{\includegraphics[width=0.33\textwidth]{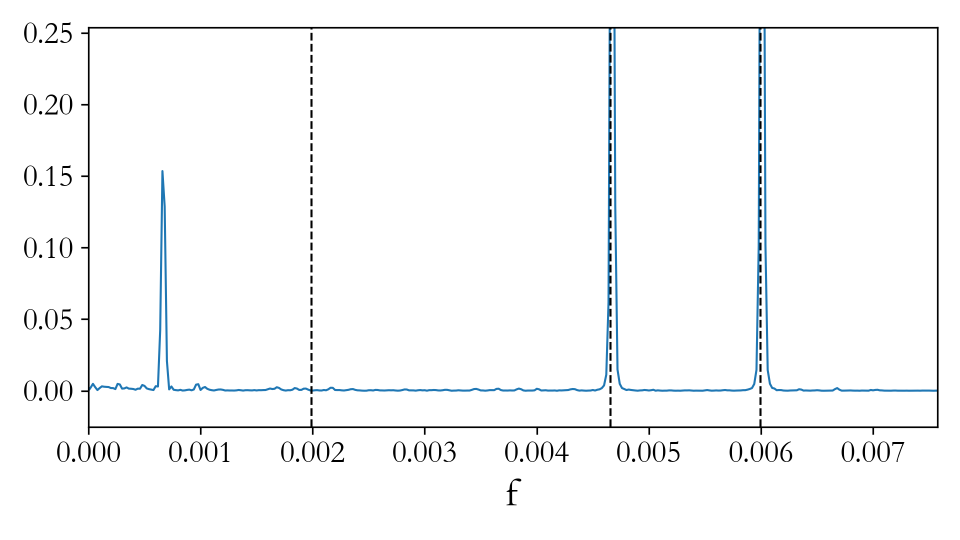}}
	\subfloat[$m=9$]{\includegraphics[width=0.33\textwidth]{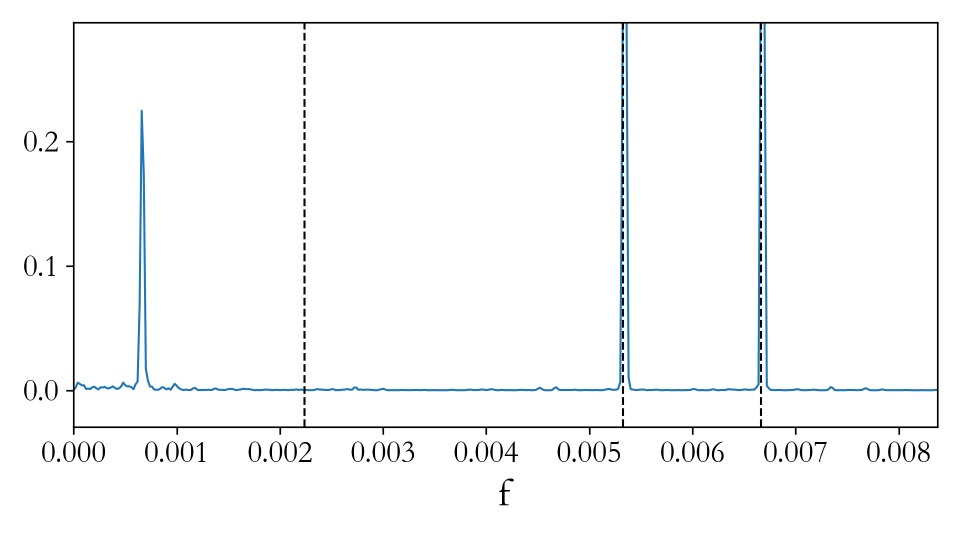}}
	\subfloat[$m=10$]{\includegraphics[width=0.33\textwidth]{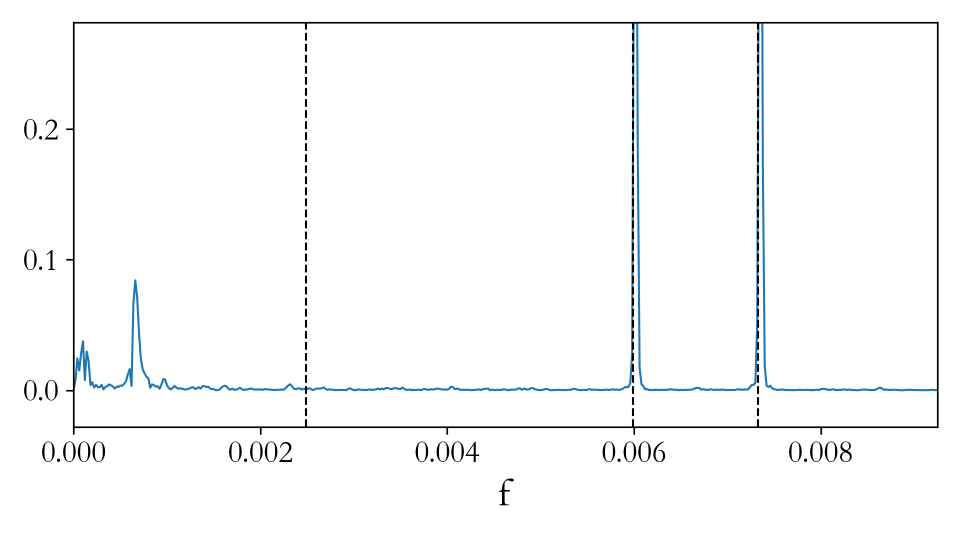}}
	\caption{Comparison between predicted (vertical dotted lines) and observed oscillation frequencies for a kinky vorton perturbed with amplitude $\epsilon=0.01$, for modes $m=2,\ldots,10$ in parameter set A. Agreement is excellent for the intermediate and higher-frequency branches. The lowest predicted frequency yields a much smaller (or absent) peak because it is dominated by longitudinal oscillations. Note that the persistent frequency at $f=0.00066$ is the ``breathing mode" of the solution and becomes more dominant for higher modes of excitation with increasing circular symmetry.}
	\label{fig:mode_osc_eval}
\end{figure*}

To test stability under non-axially symmetric perturbations, we simulate kinky vortons initialised with a multiplicatively perturbed radius. Each of the eight scalar fields (see Appendix~\ref{sec:field_reps_and_params}) is perturbed as 
\begin{equation}
\phi_i\big(r\big) \to \phi_i\big(r[1 + \epsilon\sin(m\theta)]\big)\,,
\end{equation}
where $\epsilon$ is the perturbation amplitude. For each kinky vorton in Fig.~\ref{fig:stab_vort_long}, which are predicted to be stable to all extrinsic modes (see Fig.~\ref{fig:mode_eval}), we excite modes $m=2,\ldots,10$ and evolve to $t=50000$ with $\epsilon=0.01$. Figure~\ref{fig:mode_osc_eval} shows a representative example for the $N=500$ solution in parameter set A; qualitatively identical results are obtained in other cases.

We find that the oscillation modes are well excited by our perturbations, and that the predictions for the intermediate and highest frequencies are consistent for all modes. However, we find that the smallest predicted frequency appears with a much smaller or seemingly absent peak for the higher number modes. A similar effect was observed in Ref.~\cite{Battye:2021kbd} for vortons in a $U(1)\times U(1)$ model, and was attributed to a mode that is predominantly longitudinal and, therefore, weakly coupled to radial perturbations and position diagnostics. To quantify this, we evaluate the eigenvectors of the matrix in Eq.~\eqref{eq:osc_eigen} at the predicted eigenfrequencies obtained from Eq.~\eqref{eq:freq_cubic}. The eigenvectors are normalised, and the relative magnitudes of their components determine the character of each mode: the first two components correspond to longitudinal perturbations, while the third represents the transverse perturbation. Their squared amplitudes then give the fractional longitudinal and transverse contributions.

For the kinky vorton shown in Fig.~\ref{fig:mode_osc_eval}, the lowest frequency is found to be overwhelmingly longitudinal. For example, at $m=2$ the longitudinal components exceed the transverse by approximately four orders of magnitude, increasing to six orders of magnitude by $m=10$. This lowest frequency mode, therefore, carries negligible transverse (radial) motion and is naturally suppressed in our Fourier analysis of radial oscillations. Given the agreement for all intermediate and higher frequencies, the observed dynamics are consistent with the elastic-string description.

For all kinky vortons predicted to be stable, no signs of instability or any growth of excited modes are observed over evolution times of $t=50000$. We conclude that parameter sets A and B admit dynamically stable kinky vortons in the regimes explored, and that their equilibrium radii and oscillation spectra are accurately captured by the TSA and elastic-string framework.

\subsubsection{Unstable vortons}

Having established that the elastic string description accurately captures the dynamics of the stable solutions above, we now consider configurations predicted to be unstable. 

\begin{figure}
    \centering
    \includegraphics[width=\columnwidth]{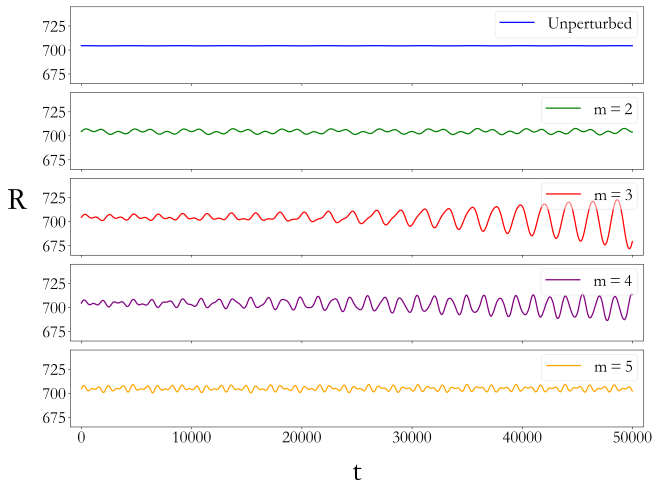}
    \caption{Evolution of the core of a domain wall at fixed $\theta=0$ during $(2+1)$-dimensional dynamics, for a kinky vorton solution in parameter set C, with $\kappa=-0.11$, $N=700$, $Q=724.5$, $R=704.5$, and $\omega=0.936$. The solution is perturbed by different external oscillation modes $m$, and the resulting dynamics are shown for each mode. Consistent with the elastic string prediction, only the $m = 3$ and $m = 4$ perturbations exhibit clear growth. All panels share identical axis scales to highlight the growing modes. Lattice sizes are $N_r = 1000$ and $N_\theta = 22144$.}
    \label{fig:unstable_vort_mode_comp}
\end{figure}

\begin{figure}
\centering
\subfloat[$t=0$]{\includegraphics[width=0.44\columnwidth]{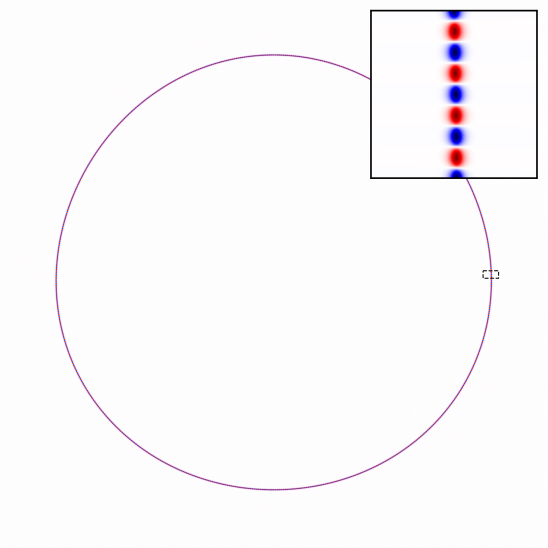}}\hspace{1pt}
\subfloat[$t=50000$]{\includegraphics[width=0.44\columnwidth]{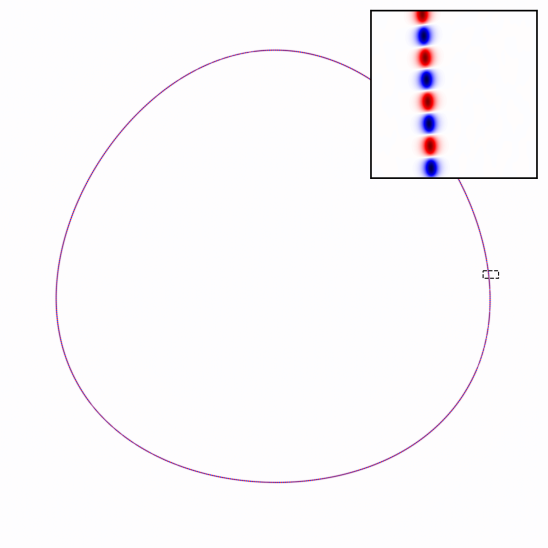}}\hspace{1pt}
\includegraphics[width=0.083\columnwidth]{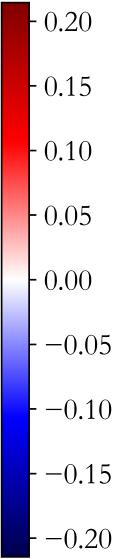}
\caption{Snapshots of the bi-linear component $R^4$ from the evolution of a kinky vorton in parameter set C, with $\kappa=-0.11$, $N=700$, $Q=724.5$, $R=704.5$, and $\omega=0.936$. Initially perturbed with an oscillation mode of $m=3$. The growth of the unstable deformation is clearly visible as the simulation progresses. Inset axes highlight the winding of the solution. Lattice sizes are $N_r = 1000$ and $N_\theta = 22144$.}
\label{fig:param_C_mode_evo}
\end{figure}

We first study a kinky vorton in parameter set C with $\kappa = -0.11$, which is predicted to be unstable to non-axially symmetric perturbations of modes $m=3$ and $m=4$ only. We perturb the solution with a range of external modes and track the resulting radial evolution, shown in Fig.~\ref{fig:unstable_vort_mode_comp}. In agreement with the elastic string prediction, only the $m=3$ and $m=4$ perturbations exhibit clear, sustained growth. All other modes of perturbation remain bounded throughout the simulations, displaying oscillations of consistent amplitude with no observable growth. Figure~\ref{fig:param_C_mode_evo} shows snapshots of the evolution of this kinky vorton for the perturbation mode $m=3$, from which one can see the clear growth of the perturbation throughout the simulation.

\begin{figure}
    \centering
    \includegraphics[width=\columnwidth]{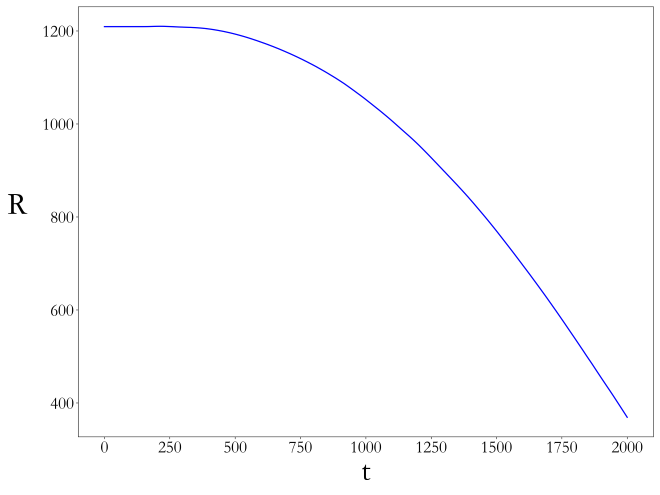}
    \caption{Evolution of the core of a domain wall at fixed $\theta=0$ during $(2+1)$-dimensional dynamics, for a kinky vorton solution in parameter set D, $\kappa = -0.12$, $N=1500$, $Q=2731.8$, $R=1209.3$, and $\omega=1.191$. The solution is predicted to be unstable to longitudinal perturbations, with the breathing mode of the vorton sourcing an instability and causing decay of the solution. Lattice sizes are $N_r = 1000$ and $N_\theta = 30816$.}
    \label{fig:unstable_param_D_rad_evo}
\end{figure}

\begin{figure}
\centering
\subfloat[$t=0$]{\includegraphics[width=0.44\columnwidth]{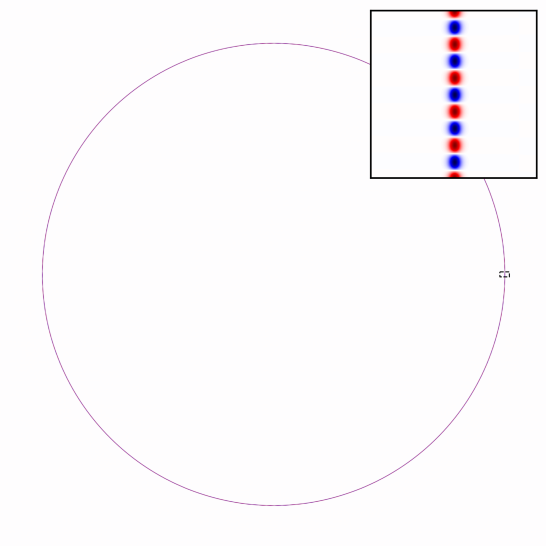}}\hspace{1pt}
\subfloat[$t=400$]{\includegraphics[width=0.44\columnwidth]{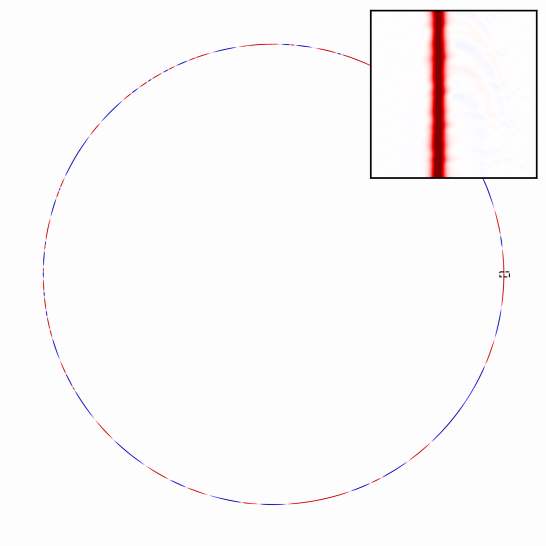}}\hspace{1pt}
\includegraphics[width=0.088\columnwidth]{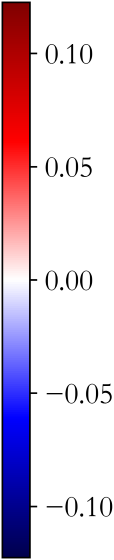}
\caption{Snapshots of the bi-linear component $R^4$ from the evolution of a kinky vorton in parameter set D, with $\kappa = -0.12$, $N=1500$, $Q=2731.8$, $R=1209.3$, and $\omega=1.191$. The solution is unstable to longitudinal perturbations, unwinding due to the onset of a pinching instability. Inset axes highlight the unwinding of the solution. Lattice sizes are $N_r = 1000$ and $N_\theta = 30816$.}
\label{fig:param_D_evo}
\end{figure}

We next consider a kinky vorton in parameter set D, with $\kappa = -0.12$, which is predicted to be unstable to longitudinal perturbations. This instability develops because $c_L$ is imaginary and in this case the associated growth rate is much larger than the growth rate due to $\Omega_{3,4}$ in parameter set C, which are suppressed by a factor of $1/R$. All other perturbation frequencies for this solution are predicted to be real but the longitudinal instability can be expected to rapidly grow, from very small initial perturbations of numerical origin. Figure~\ref{fig:unstable_param_D_rad_evo} shows that the instability indeed develops very quickly, leading to a rapid departure from the initial configuration. This behaviour is further illustrated in Fig.~\ref{fig:param_D_evo}, which shows the onset of a pinching instability, as characterized in Ref.~\cite{PhysRevD.107.063534}, and the unwinding of the solution. No longer supported by a persistent current the loop swiftly collapses under its own tension.

These instability studies further reinforce the agreement between the fully dynamical $(2+1)$-dimensional simulations and the elastic TSA description of kinky vortons in the 2HDM. Both non-axisymmetric and longitudinal pinching instabilities are accurately captured by the established frameworks in the regimes explored.

\section{Kinky Vortons in Three Dimensions}
While kinky vortons are intended as a computationally tractable proxy for three-dimensional vortons, we have also found that the vacuum topology of the $\mathbb{Z}_2$-symmetric 2HDM may admit a direct realisation of such objects in three spatial dimensions. Recent work has shown that maximally $CP$-violating domain-wall solutions exist in the model \cite{CCDW}, provided the mass parameters satisfy
\begin{equation}
	M_A < M_{H^\pm}\,, \quad M_{A}^2 \lesssim \half\left[M_H^2 +M_h^2\left(\frac{1}{f^{2}_1(0)} - 1\right) \right]\,.
\end{equation}
In this regime, a secondary, local, $CP1$ domain wall may form longitudinally upon an underlying $\mathbb{Z}_2$-symmetric wall, resulting in maximal $CP$ violation confined to the core of the secondary wall. This is topologically justified as the $\mathbb{Z}_2$ symmetry is enhanced to a $\mathbb{Z}_2\times\mathbb{Z}_2$ symmetry, with the two symmetries corresponding to flipping the sign of $R_1$ and $R_2$ respectively, but only one of them is spontaneously broken in the vacuum --- the one that corresponds to $R_1$. In the core of the wall, the spontaneously broken $\mathbb{Z}_2$ is restored and it is possible for the second $\mathbb{Z}_2$, also known as a $CP1$ symmetry, to be spontaneously broken locally. A stable two-dimensional example of this composite structure was presented in Sec.~VI of Ref.~\cite{CCDW}, and may be trivially extended to three dimensions by translational symmetry.

Here we show that the $CP1$ domain wall itself admits further non-trivial structure. Specifically, we find that localized condensates can be supported by the local $CP1$ walls. We illustrate this using the parameter set
\[
M_H = 500\,, \quad M_A = 125\,, \quad M_{H^\pm} = 250\,, \quad \tan\beta = 0.85\,,
\]
with masses given in GeV, which lies well within the mass regime which exhibits $CP$-violating domain walls.

The construction of our solution closely follows that of Ref.~\cite{CCDW}. We consider a two-dimensional spatial grid with the $\mathbb{Z}_2$ domain wall aligned along the $y$-direction and two $CP1$ domain walls localized longitudinally on the $\mathbb{Z}_2$ wall, interpolating between the two $CP$-conjugate vacua. We impose homogeneous Neumann boundary conditions in $x$, and periodic in $y$ to better approximate the composite domain wall network. In addition, we allow for a localized condensate on each $CP1$ wall in the form of an initial two-dimensional Gaussian profile. We then solve the equations of motion via gradient flow, to obtain an energy minimizing configuration. 
\begin{figure}
    \centering
    \subfloat[$R^1$]{
        \begin{minipage}[b]{0.28\columnwidth}
            \centering
            \includegraphics[width=\columnwidth]{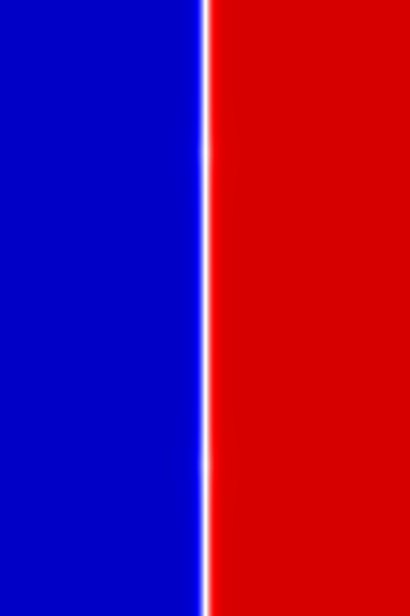}
        \end{minipage}
    }
    \subfloat[$R^2$]{
        \begin{minipage}[b]{0.28\columnwidth}
            \centering
            \includegraphics[width=\columnwidth]{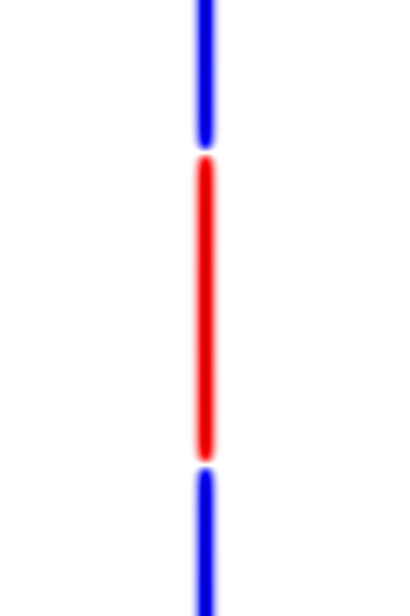}
        \end{minipage}
    }
    \subfloat[$R^\mu R_\mu$]{
        \begin{minipage}[b]{0.28\columnwidth}
            \centering
            \includegraphics[width=\columnwidth]{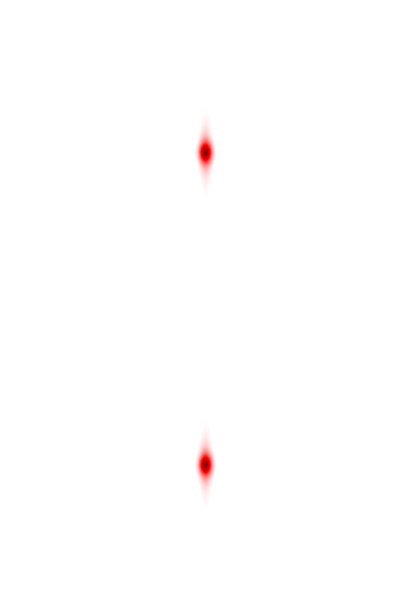}
        \end{minipage}
    \includegraphics[width=0.062\columnwidth]{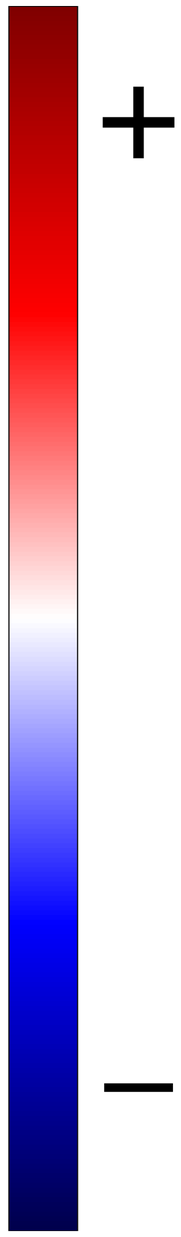}
    }\\
    \subfloat[$R^\mu$ (longitudinal)]{
        \begin{minipage}[b]{\columnwidth}
            \centering
            \includegraphics[width=\columnwidth]{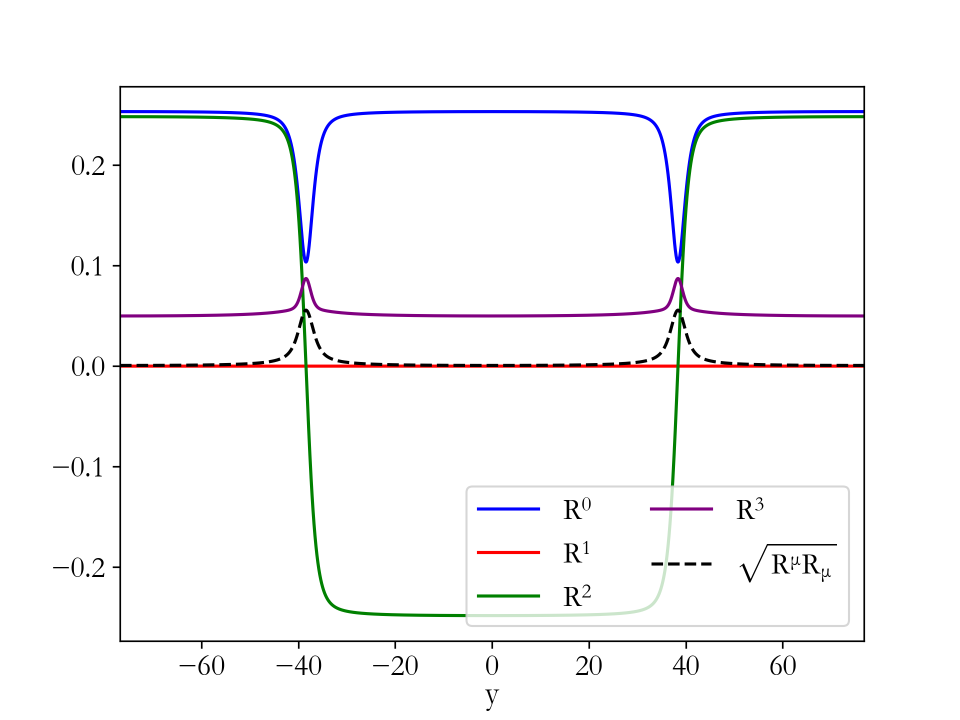}
        \end{minipage}
    }
    \caption{Stable two-dimensional composite domain-wall solution in the globally $\mathbb{Z}_2$-symmetric 2HDM. A $\mathbb{Z}_2$ domain wall supports two localized $CP1$ domain walls, each of which supports a localized condensate. We find this two-dimensional configuration to be stable, with maximal CP-violation along the primary domain wall and a localized non-zero photon mass on the secondary walls.}
    \label{fig:2d_cp_vio_condensate}
\end{figure}
The resulting solution is shown in Fig.~\ref{fig:2d_cp_vio_condensate} and exhibits a clear superposition structure: a $\mathbb{Z}_2$ domain wall supporting two longitudinal $CP1$ walls, each supporting a localized condensate. The condensate is energetically favoured on the $CP$-violating walls because at their core the $CP$ mixing angle satisfies $\xi = 0$, an analytic discussion of which can be found in Ref.~\cite{CCDW}. The existence of a condensate was not considered in the first presentation of this type of composite solution.

This structure admits a natural extension to three dimensions. In that case, the $\mathbb{Z}_2$ wall generalises to a two-dimensional sheet, on which the $CP1$ domain walls exist as one-dimensional defects. If a condensate-supporting $CP1$ wall forms a closed loop on the $\mathbb{Z}_2$ sheet, a winding of the condensate would yield kinky-vorton-like configurations. In this way, kinky vortons could be realised in three-dimensional space, but they are constrained to live on the surface of two-dimensional walls. A detailed study of their existence and stability lies beyond the scope of this work, since it would require a re-evaluation of the current-carrying ansatz and of the TSA directly in two dimensions, rather than the simpler kink solution construction of the kinky vortons we have presented. Therefore, we restrict ourselves to suggesting the existence of these composite configurations, and leave a systematic analysis to future work.

\section{Conclusions}\label{sec:conclusions}
The primary objective of this work has been to provide an initial stepping stone for the construction of vortons in the two-Higgs-Doublet Model. We have delivered this by demonstrating the existence of two-dimensional current-carrying ring solutions -- kinky vortons -- in the $\mathbb{Z}_2$-symmetric variant of the global model. In the restricted parameter regimes relevant to these solutions, the effective equivalence of the $\mathbb{Z}_2$ and $U(1)$-symmetric potentials implies that kinky vortons provide a suitable and computationally tractable proxy for vortons in the $U(1)$-symmetric 2HDM.

We have presented multiple examples of dynamically stable kinky vortons in the model, shown to persist under non-axially symmetric perturbations over long timescales. Crucially, we demonstrated that the properties of these solutions, winding number, charge, and equilibrium radii, are well described by the thin-string approximation, with their oscillation spectra and stability to extrinsic oscillation modes accurately captured by the elastic string formalism. The successful application of these established frameworks to vorton-like defects in the 2HDM provides strong evidence that vorton solutions can be constructed consistently within this extension of the Standard Model.

Although the solutions presented here are seemingly restricted to the large-radius limit and to the magnetic regime of the current-carrying ansatz, these restrictions are expected to be ameliorated in the gauged theory. Subsequent studies will focus on constructing both vortons and kinky vortons in the gauged versions of the model, in order to establish the suitability of kinky vortons as a robust computationally simpler proxy for vortons in the full theory, which would allow for a greater analysis of the parameter space for observational constraint purposes.

We have also extended the class of non-trivial defects known to arise within the model, demonstrating that the $\mathbb{Z}_2$-symmetric theory may naturally admit kinky-vorton-like defects even in a three dimensional setting. Given the phenomenological relevance of the $\mathbb{Z}_2$-symmetric 2HDM, the possible formation of superconducting loops on maximally $CP$-violating domain walls during the electroweak phase transition opens further avenues of study. Such objects could lead to an abundance of heavy relics of the electroweak phase transition, potentially relevant to open problems such as the origin of the matter–antimatter asymmetry of the Universe. These possibilities are, of course, purely speculative and are left for future work involving particle interactions with the domain-wall solutions described here and in Ref.~\cite{CCDW}.

Beyond a cosmological setting, current-carrying vortex rings have previously been studied in two-component Bose--Einstein condensates as laboratory analogues of cosmic vortons \cite{Metlitski_2004, PhysRevA.85.053639}. In particular, it has been shown that stable vortex loops may be supported by localized condensates and persistent currents in such models. Our presentation of stable kinky vortons in the 2HDM suggests that vorton-like structures could arise in more complicated multi-component condensed matter systems.

\appendix
\section{Representations of the 2HDM}\label{sec:field_reps_and_params}

In this work we make use of multiple equivalent representations of the fields in the global two-Higgs-Doublet Model, each chosen according to its suitability for either numerical or analytic applications.

The linear representation,
\begin{equation}
\Phi_1 = \begin{pmatrix} \phi_1 + i\phi_2 \\ \phi_3 + i\phi_4 \end{pmatrix}\,, \quad \Phi_2 = \begin{pmatrix} \phi_5 + i\phi_6 \\ \phi_7 + i\phi_8 \end{pmatrix}\,,
\end{equation}
treats the two scalar doublets as eight real fields. This form is used for all dynamical simulations, as the resulting equations of motion exhibit simple symmetry relations under exchanges of individual field components, enabling an efficient lattice code implementation.

The general representation, used for example in Refs.~\cite{Battye2021SDW, BATTYE2025139311, CCDW}, expresses a generic field configuration as a state, with a reduced number of degrees of freedom, acted upon by a general $U \in U(1)\times SU(2)_L$ transformation,
\begin{equation}
\Phi = \begin{pmatrix} \Phi_1 \\ \Phi_2 \end{pmatrix} = \frac{v_{\rm SM}}{\sqrt{2}}(\sigma^0 \times U)\begin{pmatrix} 0 \\ f_1 \\ f_+ \\ f_2e^{i\xi} \end{pmatrix}\,,
\end{equation}
where $\sigma^0$ is the $2 \times 2$ identity. This representation is particularly useful for constructing solution ans{\"a}tze as the field components offer interpretable forms, such as condensates in $f_+$ result in a non-zero photon mass, and simple analytic interpretations of field component couplings in the equations of motion.

Topological defects are most naturally described in the bi-linear field space formalism, introduced in Refs.~\cite{Nishi:2006tg, Ivanov:2006yq, Maniatis_2006}, where the potential is described fully using the four-vector
\begin{equation}
R^\mu = \Phi^\dagger(\sigma^\mu \otimes \sigma^0)\Phi  = \begin{pmatrix} {\Phi_1^\dagger \Phi_1 + \Phi_2^\dagger \Phi_2} \\ {\Phi_1^\dagger \Phi_2 + \Phi_2^\dagger \Phi_1} \\ {-i[\Phi_1^\dagger \Phi_2 - \Phi_2^\dagger \Phi_1]} \\ {\Phi_1^\dagger \Phi_1 - \Phi_2^\dagger \Phi_2} \end{pmatrix}\,,
\label{eq:bi-linear_4_vec}
\end{equation}
such that the potential takes the form
\begin{equation}
    V = -\tfrac{1}{2}M_\mu R^\mu + \tfrac{1}{4}L_{\mu \nu} R^\mu R^\nu\,.
\end{equation}
The coefficients $M_\mu$ and $L_{\mu\nu}$ encode the quadratic mass parameters $\mu_1^2$, $\mu_2^2$, $m_{12}^2$ and the quartic couplings $\lambda_{1,\dots,7}$; explicit expressions may be found in Ref.~\cite{Battye2011VT}. Charged degrees of freedom can be incorporated through the $SU(2)_L$-invariant $\Phi_1^T i\sigma^2 \Phi_2$, promoting $R^\mu$ to the null six-vector $R^A$, as shown in Eq.~\eqref{eq:bi-linear_6_vec}.

The potential of the 2HDM, which can also be represented in the bi-linear formalism as detailed above, takes the general form
\begin{align}
	V = & -\mu_{1}^2 (\Phi_1^\dag \Phi_1) - \mu_2^2 (\Phi_2^\dag \Phi_2) -m_{12}^2(\Phi_1^\dag \Phi_2) -m_{12}^{*2}(\Phi_2^\dag \Phi_1)\cr
	& + \lambda_1 (\Phi_1^\dag \Phi_1)^2 + \lambda_2 (\Phi_2^\dag \Phi_2)^2  + \lambda_3 (\Phi_1^\dag \Phi_1)(\Phi_2^\dag \Phi_2) \cr 
	& + \lambda_4 (\Phi_1^\dag \Phi_2)(\Phi_2^\dag \Phi_1) + \tfrac{1}{2} \lambda_5 (\Phi_1^\dagger \Phi_2)^2 + \tfrac{1}{2} \lambda_5^* (\Phi_2^\dagger \Phi_1)^2 \cr
	& + \lambda_6(\Phi_1^\dag \Phi_1)(\Phi_1^\dag \Phi_2) + \lambda_6^*(\Phi_1^\dag \Phi_1)(\Phi_2^\dag \Phi_1)\cr
	& + \lambda_7(\Phi_2^\dag \Phi_2)(\Phi_1^\dag \Phi_2) + \lambda_7^*(\Phi_2^\dag \Phi_2)(\Phi_2^\dag \Phi_1)\,,
\label{eq:2HDM_Potential}
\end{align}
where a $\mathbb{Z}_2$-symmetric potential can be obtained by imposing
\begin{equation}
m_{12}^2 = \lambda_6 = \lambda_7 = 0, \quad \lambda_5 \in \Re\,,
\label{eq:Z2_restrictions}
\end{equation}
and a global $U(1)$ symmetry follows upon further setting $\lambda_5=0$.

An analysis of the mass matrices, see Ref.~\cite{Battye2021SDW}, allows the parameters of the $\mathbb{Z}_2$-symmetric potential to be expressed in terms of the physical scalar masses, the SM VEV, $v_{\rm SM}$, and the mixing angles of the CP-even and CP-odd fields, $\alpha$ and $\beta$, 
\begin{align}
\mu_1^2 & = \tfrac{1}{2}\left[M_h^2 c_\alpha^2+ M_H^2 s_\alpha^2+ \left( M_h^2 - M_H^2 \right) c_\alpha s_\alpha \tan\beta\right]\,,\cr
\mu_2^2 & = \tfrac{1}{2}\left[M_h^2 s_\alpha^2+ M_H^2 c_\alpha^2+ \left( M_h^2 - M_H^2 \right) c_\alpha s_\alpha \cot\beta\right]\,,\cr
\lambda_1 & = \frac{M_h^2 c_\alpha^2 + M_H^2 s_\alpha^2}{2v_{\rm SM}^2 c_\beta^2 }\,, \qquad
\lambda_2 = \frac{M_h^2 s_\alpha^2 + M_H^2 c_\alpha^2}{2v_{\rm SM}^2 s_\beta^2 }\,, \cr
\lambda_3 & = \frac{(M_h^2 - M_H^2)\, c_\alpha s_\alpha + 2 M_{H^\pm}^2\, c_\beta s_\beta}{v_{\rm SM}^2c_\beta s_\beta }\,,\cr
\lambda_4 & = \frac{M_A^2 - 2M^2_{H^\pm}}{v_{\rm{SM}}^2}\,, \qquad |\lambda_5| = \frac{M_A^2}{v_{\rm{SM}}^2}\,,
\label{eq:mass_param_rel}
\end{align}
where $s_x \equiv \sin x$ and $c_x \equiv \cos x$. 

Within the alignment limit, where $\cos (\alpha - \beta) = 1$, imposing the parameter restrictions of Eq.~(\ref{eq:Z2_restrictions}) and substituting Eq.~(\ref{eq:mass_param_rel}) into Eq.~(\ref{eq:2HDM_Potential}) yields the $\mathbb{Z}_2$-symmetric potential used throughout this work, given explicitly in Eq.~(\ref{eq:Z2_sym_pot}).

\section{Numerical Techniques}\label{sec:numericals}

All numerical work presented in this paper uses units of length $M_h^{-1}$ and energy densities $v^2_{\rm SM}M_h^2$.

\subsection{Kink solutions}\label{sec:numericals-kinks}
The kink solutions used to construct kinky vortons are obtained by numerically solving the equations of motion associated with the superconducting ansatz of Eq.~(\ref{eq:supercond_ansatz}). The fields are discretised on a one-dimensional lattice of $n_x$ points with uniform spacing $\Delta x$. Spatial derivatives are approximated using fourth-order finite-difference stencils, and homogeneous Neumann boundary conditions are imposed to obtain energy-minimising configurations. Initial field profiles are chosen as approximate analytic forms motivated by the known superconducting solutions of Ref.~\cite{CCDW}.

The static equations of motion are
\begin{align}
& \partial_x^2 g_1 - g_1\!\left(-\mu_1^2+ \lambda_1 (g_1^2 + g_2^2) + \tfrac{1}{2}\lambda_3 (g_3^2 + g_4^2) - \kappa\right)\cr
& - \tfrac{1}{2}(\lambda_4 - \lambda_5)\, g_3 \,(g_1 g_3 + g_2 g_4) \equiv F_{g_1} = 0\,,
\end{align}
\begin{align}
&\partial_x^2 g_2- g_2\!\left(-\mu_1^2+ \lambda_1 (g_1^2 + g_2^2)+ \tfrac{1}{2}\lambda_3 (g_3^2 + g_4^2)\right)\cr
& - \tfrac{1}{2}(\lambda_4 - \lambda_5)\, g_4 \,(g_1 g_3 + g_2 g_4) \equiv F_{g_2} = 0\,,
\end{align}
\begin{align}
&\partial_x^2 g_3- g_3\!\left(-\mu_2^2+ \lambda_2 (g_3^2 + g_4^2)+ \tfrac{1}{2}\lambda_3 (g_1^2 + g_2^2)- \kappa\right) \cr
&- \tfrac{1}{2}(\lambda_4 - \lambda_5)\, g_1 \,(g_1 g_3 + g_2 g_4) \equiv F_{g_3} = 0\,,
\end{align}
\begin{align}
&\partial_x^2 g_4- g_4\!\left(-\mu_2^2+ \lambda_2 (g_3^2 + g_4^2)+ \tfrac{1}{2}\lambda_3 (g_1^2 + g_2^2)\right) \cr
& - \tfrac{1}{2}(\lambda_4 - \lambda_5)\, g_2 \,(g_1 g_3 + g_2 g_4 \equiv F_{g_4}) = 0\,.
\end{align}
We use a damped Newton–Raphson relaxation scheme, where at each iteration step, $n$, the fields are updated pointwise according to
\begin{equation}
g_i^{\,n+1}=g_i^{\,n}- w \,\frac{F_{g_i}^{\,n}}{\partial F_{g_i}^{\,n} / \partial g_i}\,,
\end{equation}
where the denominator corresponds to a local (diagonal) approximation of the Jacobian. The relaxation parameter is chosen as $w = 0.75$, found to provide stable convergence. Iteration proceeds until all field updates fall below a predefined tolerance of $\delta = 10^{-7}$.

\subsection{Gradient flow of kinky vorton solutions}\label{sec:numericals-GF}
Static, circularly symmetric, kinky vorton solutions with fixed winding number $N$ and charge $Q$ are obtained using a gradient flow relaxation scheme in polar coordinates, performed on an annulus to avoid the coordinate singularity at the centre. We then exploit the rotational symmetry of kinky vorton solutions to reduce the problem to the evolution of a single radial profile while retaining the full two-dimensional structure through analytic treatment of angular and temporal derivatives. 

From the polar-coordinate ansatz of Eq. (\ref{eq:CC_DW}) we obtain a coupled system of radial equations of motion. For the charged components, $g_1$ and $g_3$ the equations take the form
\begin{equation}
\partial_r^2 g_i + \frac{1}{r}\,\partial_r g_i - \frac{N^2}{r^2}\, g_i + \omega^2\, g_i - \frac{\partial V}{\partial g_i} \equiv F_{g_i}= 0\,,
\end{equation}
with analogous equations for the neutral components, $g_2$ and $g_4$, which lack the angular and frequency-dependent terms. To obtain static solutions, the radial fields are evolved in a fictitious time $\tau$ via gradient flow,
\begin{equation}
\frac{\partial g_i(r,\tau)}{\partial \tau} = F_{g_i}(r,\tau)\,.
\end{equation}

A single radial profile is discretised on a one-dimensional lattice of $n_r$ points with uniform spacing $\Delta r$, with spatial derivatives once again approximated to fourth order. Evolution in $\tau$ is performed using an explicit first-order finite difference update,
\begin{equation}
g_i^{\,n+1}(r) = g_i^{\,n}(r)+ \Delta \tau \, F_{g_i}^{\,n}(r)\,,
\end{equation}
with the fictitious time-step, $\Delta \tau$, chosen proportional to ${\Delta r}^2$ to ensure numerical stability. 

Neumann boundary conditions are imposed at both radial boundaries, and an initial field configuration is constructed by placing a one-dimensional kink solution, as described in Appendix.~\ref{sec:numericals-kinks}, at a predicted kinky vorton radius $R_*$ for a given winding number $N$, as determined by the thin string approximation. The conserved charge $Q$ is maintained dynamically during the evolution; at each iteration, the charge integral, $\Sigma_2 = \int r dr (g_1^2 + g_3^2)$, is evaluated and the frequency  $\omega$ is rescaled to maintain $Q = 2\pi \Sigma_2 \omega$. Gradient flow continues until all field updates are again below a tolerance of $\delta = 10^{-7}$. The resulting configurations correspond to radial profiles of circularly symmetric kinky vorton solutions, for the specified $N$ and $Q$, with the radii $R$ and frequency $\omega$ converged dynamically during the evolution.

\subsection{Full dynamical evolution}\label{sec:numericals-sims}

Simulations of kinky vorton solutions are performed using polar coordinates. The scalar fields, in the linear representation, are evolved on a discretised annulus of $N_r$ grid points in the radial direction and $N_\theta$ in the angular direction, with uniform spacings $\Delta r$ and $\Delta \theta = 2\pi / N_\theta$, respectively. Periodic boundary conditions are imposed in the angular direction at $\theta = 0 \equiv 2\pi$, while Neumann boundary conditions are used at both radial boundaries. Spatial derivatives are approximated to fourth order and temporal derivatives to second order, with time evolution carried out using a standard leapfrog scheme. 

A fixed radial spacing of $\Delta r = 0.20$ is used throughout, with $N_r$ chosen to accommodate the spatial extent of each solution. The annulus is centred on the loop, with its inner and outer radial boundaries placed at distances at least an order of magnitude greater than the characteristic width of the solution. For each simulation, $N_\theta$ is set such that $r\Delta\theta \approx \Delta r = 0.20$ along the domain wall loop, ensuring comparable numerical accuracy in all directions around the solution.

This polar approach yields identical dynamical results to those obtained using Cartesian-grid simulations, while being significantly more efficient for the large-radius configurations studied here. The total number of grid points required to resolve the solution is greatly reduced compared to a Cartesian grid large enough to contain the same configuration. This allows solutions to be evolved over substantially longer timescales, enabling a more robust analysis of the stability of kinky vorton solutions. All numerical results were verified to be insensitive to modest variations in resolution and annulus size.

\bibliography{References}

@book{Vilenkin278400,
      author        = "Vilenkin, Alexander and Shellard, E Paul S",
      title         = "{Cosmic strings and other topological defects}",
      publisher     = "Cambridge Univ. Press",
      address       = "Cambridge",
      series        = "Cambridge monographs on mathematical physics",
      year          = "1994",
}

@article{Battye2008KV,
	doi = {10.1016/j.nuclphysb.2008.07.034}, 
	year = 2008,
	month = {dec},
	publisher = {Elsevier {BV}},
	volume = {805},
	number = {1-2},
	pages = {287--304},
	author = {Richard A. Battye and Paul M. Sutcliffe},
	title = {Kinky Vortons},
	journal = {Nuclear Physics B}
}

@article{Battye2009SKV,
	doi = {10.1103/physrevd.80.085024},
	year = 2009,
	month = {oct},
	publisher = {American Physical Society ({APS})},
	volume = {80},
	number = {8},
	author = {Richard A. Battye and Paul M. Sutcliffe},
	title = {Stability and the equation of state for kinky vortons},
	journal = {Physical Review D}
}

@article{PILAFTSIS2012465,
title = {On the classification of accidental symmetries of the two Higgs doublet model potential},
journal = {Physics Letters B},
volume = {706},
number = {4},
pages = {465-469},
year = {2012},
issn = {0370-2693},
doi = {https://doi.org/10.1016/j.physletb.2011.11.047},
url = {https://www.sciencedirect.com/science/article/pii/S0370269311014262},
author = {Apostolos Pilaftsis}
}

@article{Battye2011VT,
	doi = {10.1007/jhep08(2011)020},
	year = 2011,
	month = {aug}, 
	publisher = {Springer Science and Business Media {LLC}},
	volume = {2011},
	number = {8},
	author = {Richard A. Battye and Gary D. Brawn and Apostolos Pilaftsis},
	title = {Vacuum topology of the two Higgs doublet model},
	journal = {Journal of High Energy Physics}
}

@article{Battye2021SDW,
	doi = {10.1007/jhep01(2021)105}, 
	year = 2021,
	month = {jan},
	publisher = {Springer Science and Business Media {LLC}},
	volume = {2021},
	number = {1},
	author = {Richard A. Battye and Apostolos Pilaftsis and Dominic G. Viatic},
	title = {Simulations of domain walls in Two Higgs Doublet Models},
	journal = {Journal of High Energy Physics}
}

@article{PhysRevD.105.056007,
  title = {Charged and $CP$-violating kink solutions in the two-Higgs-doublet model},
  author = {Law, Kai Hong and Pilaftsis, Apostolos},
  journal = {Phys. Rev. D},
  volume = {105},
  issue = {5},
  pages = {056007},
  numpages = {14},
  year = {2022},
  month = {Mar},
  publisher = {American Physical Society},
  doi = {10.1103/PhysRevD.105.056007},
  url = {https://link.aps.org/doi/10.1103/PhysRevD.105.056007}
}

@article{Branco_2012,
   title={Theory and phenomenology of two-Higgs-doublet models},
   volume={516},
   ISSN={0370-1573},
   DOI={10.1016/j.physrep.2012.02.002},
   number={1–2},
   journal={Physics Reports},
   publisher={Elsevier BV},
   author={Branco, G.C. and Ferreira, P.M. and Lavoura, L. and Rebelo, M.N. and Sher, Marc and Silva, João P.},
   year={2012},
   month=jul, pages={1–102} 
}

@article{Nishi:2006tg,
  title = {$CP$ violation conditions in $N$-Higgs-doublet potentials},
  author = {Nishi, C. C.},
  journal = {Phys. Rev. D},
  volume = {74},
  issue = {3},
  pages = {036003},
  numpages = {16},
  year = {2006},
  month = {Aug},
  publisher = {American Physical Society},
  doi = {10.1103/PhysRevD.74.036003},
  url = {https://link.aps.org/doi/10.1103/PhysRevD.74.036003}
}

@article{Ivanov:2006yq,
  title = {Minkowski space structure of the Higgs potential in the two-Higgs-doublet model},
  author = {Ivanov, I. P.},
  journal = {Phys. Rev. D},
  volume = {75},
  issue = {3},
  pages = {035001},
  numpages = {20},
  year = {2007},
  month = {Feb},
  publisher = {American Physical Society},
  doi = {10.1103/PhysRevD.75.035001},
  url = {https://link.aps.org/doi/10.1103/PhysRevD.75.035001}
}

@article{PhysRevD.107.063534,
  title = {Pinching instabilities in superconducting cosmic strings},
  author = {Battye, R. A. and Cotterill, S. J.},
  journal = {Phys. Rev. D},
  volume = {107},
  issue = {6},
  pages = {063534},
  numpages = {14},
  year = {2023},
  month = {Mar},
  publisher = {American Physical Society},
  doi = {10.1103/PhysRevD.107.063534},
  url = {https://link.aps.org/doi/10.1103/PhysRevD.107.063534}
}

@article{Ivanov_2008,
   title={Minkowski space structure of the Higgs potential in the two-Higgs-doublet model. II. Minima, symmetries, and topology},
   volume={77},
   ISSN={1550-2368},
   DOI={10.1103/physrevd.77.015017},
   number={1},
   journal={Physical Review D},
   publisher={American Physical Society (APS)},
   author={Ivanov, I. P.},
   year={2008},
   month=jan 
}

@article{Maniatis_2006,
   title={Stability and symmetry breaking in the general two-Higgs-doublet model},
   volume={48},
   ISSN={1434-6052},
   DOI={10.1140/epjc/s10052-006-0016-6},
   number={3},
   journal={The European Physical Journal C},
   publisher={Springer Science and Business Media LLC},
   author={Maniatis, M. and von Manteuffel, A. and Nachtmann, O. and Nagel, F.},
   year={2006},
   month=oct, pages={805–823}
}

@article{Witten:1984eb,
    author = "Witten, Edward",
    title = "{Superconducting Strings}",
    reportNumber = "PRINT-84-0763 (IAS,-PRINCETON)",
    doi = "10.1016/0550-3213(85)90022-7",
    journal = "Nucl. Phys. B",
    volume = "249",
    pages = "557--592",
    year = "1985"
}

@article{DAVIS1989209,
	title = {Cosmic vortons},
	journal = {Nuclear Physics B},
	volume = {323},
	number = {1},
	pages = {209-224},
	year = {1989},
	issn = {0550-3213},
	doi = {https://doi.org/10.1016/0550-3213(89)90594-4},
	author = {R.L. Davis and E.P.S. Shellard},
}

@article{Battye_2023,
   title={Global monopoles in the two-Higgs-doublet-model},
   volume={844},
   ISSN={0370-2693},
   DOI={10.1016/j.physletb.2023.138091},
   journal={Physics Letters B},
   publisher={Elsevier BV},
   author={Battye, Richard A. and Cotterill, Steven J. and Viatic, Dominic G.},
   year={2023},
   month=sep, pages={138091}
}

@article{PhysRevD_8_1226,
  title = {A Theory of Spontaneous T Violation},
  author = {Lee, T. D.},
  journal = {Phys. Rev. D},
  volume = {8},
  issue = {4},
  pages = {1226--1239},
  numpages = {0},
  year = {1973},
  month = {Aug},
  publisher = {American Physical Society},
  doi = {10.1103/PhysRevD.8.1226}
}

@article{Keus_2016,
   title={CP violating Two-Higgs-Doublet Model: constraints and LHC predictions},
   volume={2016},
   ISSN={1029-8479},
   DOI={10.1007/jhep04(2016)048},
   number={4},
   journal={Journal of High Energy Physics},
   publisher={Springer Science and Business Media LLC},
   author={Keus, Venus and King, Stephen F. and Moretti, Stefano and Yagyu, Kei},
   year={2016},
   month=apr, pages={1–24} 
}

@article{Dorsch_2017,
   title={A second Higgs doublet in the early universe: baryogenesis and gravitational waves},
   volume={2017},
   ISSN={1475-7516},
   DOI={10.1088/1475-7516/2017/05/052},
   number={05},
   journal={Journal of Cosmology and Astroparticle Physics},
   publisher={IOP Publishing},
   author={Dorsch, G.C. and Huber, S.J. and Konstandin, T. and No, J.M.},
   year={2017},
   month=may, pages={052–052} 
}

@article{Fromme2006,
   title={Baryogenesis in the two-Higgs doublet model},
   volume={2006},
   ISSN={1029-8479},
   DOI={10.1088/1126-6708/2006/11/038},
   number={11},
   journal={Journal of High Energy Physics},
   publisher={Springer Science and Business Media LLC},
   author={Fromme, Lars and Huber, Stephan J and Seniuch, Michael},
   year={2006},
   month=nov, pages={038–038} 
}

@article{Grzadkowski_2009,
   title={Natural multi-Higgs model with dark matter and CP violation},
   volume={80},
   ISSN={1550-2368},
   DOI={10.1103/physrevd.80.055013},
   number={5},
   journal={Physical Review D},
   publisher={American Physical Society (APS)},
   author={Grzadkowski, B. and Ogreid, O. M. and Osland, P.},
   year={2009},
   month=sep 
}

@article{Garaud_2013,
   title={Stable Cosmic Vortons},
   volume={111},
   ISSN={1079-7114},
   DOI={10.1103/physrevlett.111.171602},
   number={17},
   journal={Physical Review Letters},
   publisher={American Physical Society (APS)},
   author={Garaud, Julien and Radu, Eugen and Volkov, Mikhail S.},
   year={2013},
   month=oct 
}

@article{Workman:2022ynf,
    author = "Workman, R. L. and Others",
    collaboration = "Particle Data Group",
    title = "{Review of Particle Physics}",
    doi = "10.1093/ptep/ptac097",
    journal = "PTEP",
    volume = "2022",
    pages = "083C01",
    year = "2022"
}

@article{Arbey_2018,
   title={Status of the charged Higgs boson in two Higgs doublet models},
   volume={78},
   ISSN={1434-6052},
   url={http://dx.doi.org/10.1140/epjc/s10052-018-5651-1},
   DOI={10.1140/epjc/s10052-018-5651-1},
   number={3},
   journal={The European Physical Journal C},
   publisher={Springer Science and Business Media LLC},
   author={Arbey, A. and Mahmoudi, F. and Stål, O. and Stefaniak, T.},
   year={2018},
   month=mar 
}

@article{BATTYE2025139311,
title = {Percolation of domain walls in the two-Higgs doublet model},
journal = {Physics Letters B},
volume = {862},
pages = {139311},
year = {2025},
issn = {0370-2693},
doi = {https://doi.org/10.1016/j.physletb.2025.139311},
url = {https://www.sciencedirect.com/science/article/pii/S0370269325000711},
author = {Richard A. Battye and Steven J. Cotterill and Eva Sabater Andres and Adam K. Thomasson}
}

@article{Battye:2024dvw,
    author = "Battye, Richard and Cotterill, Steven",
    title = "{Superconducting strings in the two-Higgs doublet model}",
    eprint = "2410.03300",
    archivePrefix = "arXiv",
    journal = "Physics Letters B",
    primaryClass = "hep-ph",
    year = "2024"
}

@article{Battye:2024iec,
    author = "Battye, R. A. and Cotterill, S. J.",
    title = "{Spontaneous Hopf Fibration in the Two-Higgs-Doublet Model}",
    eprint = "2401.06567",
    archivePrefix = "arXiv",
    primaryClass = "hep-ph",
    doi = "10.1103/PhysRevLett.132.061601",
    journal = "Phys. Rev. Lett.",
    volume = "132",
    number = "6",
    pages = "061601",
    year = "2024"
}

@article{Battye:2021kbd,
    author = "Battye, Richard A. and Cotterill, Steven J. and Pearson, Jonathan A.",
    title = "{A detailed study of the stability of vortons}",
    eprint = "2112.08066",
    archivePrefix = "arXiv",
    primaryClass = "hep-ph",
    doi = "10.1007/JHEP04(2022)005",
    journal = "JHEP",
    volume = "04",
    pages = "005",
    year = "2022"
}

@article{Battye:2021sji,
    author = "Battye, R. A. and Cotterill, S. J.",
    title = "{Stable Cosmic Vortons in Bosonic Field Theory}",
    eprint = "2111.07822",
    archivePrefix = "arXiv",
    primaryClass = "hep-ph",
    doi = "10.1103/PhysRevLett.127.241601",
    journal = "Phys. Rev. Lett.",
    volume = "127",
    number = "24",
    pages = "241601",
    year = "2021"
}

@article{Battye:2008mm,
    author = "Battye, Richard A. and Sutcliffe, Paul M.",
    title = "{Vorton construction and dynamics}",
    eprint = "0812.3239",
    archivePrefix = "arXiv",
    primaryClass = "hep-th",
    reportNumber = "DCPT-08-69",
    doi = "10.1016/j.nuclphysb.2009.01.021",
    journal = "Nucl. Phys. B",
    volume = "814",
    pages = "180--194",
    year = "2009"
}

@article{Lemperiere:2003yt,
    author = "Lemperiere, Y. and Shellard, E. P. S.",
    title = "{Vorton existence and stability}",
    eprint = "hep-ph/0305156",
    archivePrefix = "arXiv",
    doi = "10.1103/PhysRevLett.91.141601",
    journal = "Phys. Rev. Lett.",
    volume = "91",
    pages = "141601",
    year = "2003"
}

@article{LEMPERIERE2003511,
title = {On the behaviour and stability of superconducting currents},
journal = {Nuclear Physics B},
volume = {649},
number = {3},
pages = {511-525},
year = {2003},
issn = {0550-3213},
doi = {https://doi.org/10.1016/S0550-3213(02)01028-3},
url = {https://www.sciencedirect.com/science/article/pii/S0550321302010283},
author = {Y Lemperiere and E.P.S Shellard}
}

@article{CCDW,
      title = {Complete classification of domain wall solutions in the $\mathbb{Z}_2$-symmetric 2HDM},
  author = {Battye, Richard A. and Cotterill, Steven J. and Thomasson, Adam K.},
  journal = {Phys. Rev. D},
  volume = {112},
  issue = {11},
  pages = {115015},
  numpages = {26},
  year = {2025},
  month = {Dec},
  publisher = {American Physical Society},
  doi = {10.1103/p1sq-38cc},
  url = {https://link.aps.org/doi/10.1103/p1sq-38cc}
}

@article{CARTER1993151,
title = {Dynamic Instability Criterion for Circular String Loops},
journal = {Annals of Physics},
volume = {227},
number = {1},
pages = {151-171},
year = {1993},
issn = {0003-4916},
doi = {https://doi.org/10.1006/aphy.1993.1078},
url = {https://www.sciencedirect.com/science/article/pii/S000349168371078X},
author = {B. Carter and X. Martin}
}

@article{ETO2018447,
title = {Constraints on two Higgs doublet models from domain walls},
journal = {Physics Letters B},
volume = {785},
pages = {447-453},
year = {2018},
issn = {0370-2693},
doi = {https://doi.org/10.1016/j.physletb.2018.09.002},
url = {https://www.sciencedirect.com/science/article/pii/S0370269318306944},
author = {Minoru Eto and Masafumi Kurachi and Muneto Nitta},
}

@article{Metlitski_2004,
doi = {10.1088/1126-6708/2004/06/017},
url = {https://doi.org/10.1088/1126-6708/2004/06/017},
year = {2004},
month = {jun},
publisher = {},
volume = {2004},
number = {06},
pages = {017},
author = {Max A. Metlitski and Ariel R. Zhitnitsky},
title = {Vortons in two component Bose-Einstein condensates},
journal = {Journal of High Energy Physics}
}

@article{PhysRevA.85.053639,
  title = {Creating vortons and three-dimensional skyrmions from domain-wall annihilation with stretched vortices in Bose-Einstein condensates},
  author = {Nitta, Muneto and Kasamatsu, Kenichi and Tsubota, Makoto and Takeuchi, Hiromitsu},
  journal = {Phys. Rev. A},
  volume = {85},
  issue = {5},
  pages = {053639},
  numpages = {11},
  year = {2012},
  month = {May},
  publisher = {American Physical Society},
  doi = {10.1103/PhysRevA.85.053639},
  url = {https://link.aps.org/doi/10.1103/PhysRevA.85.053639}
}

\end{document}